\documentclass[final,5p,twocolumn,times,numbers,sort&compress]{elsarticle}

\usepackage{amssymb}
\usepackage[fleqn]{amsmath}
\usepackage{algorithm}
\usepackage{algpseudocode}
\usepackage{graphicx}
\usepackage{booktabs}
\usepackage{siunitx}
\usepackage{hyperref}
\usepackage{float}
\usepackage{enumitem}
\usepackage{multirow}
\usepackage{multicol} 
\usepackage{nomencl}
\usepackage{framed}
\usepackage{booktabs}
\makenomenclature 
\renewcommand*\nompreamble{%
  \begingroup
  \small
  \setlength{\columnsep}{8pt}
  \setlength{\multicolsep}{2pt}
  \begin{multicols}{2}
}

\renewcommand*\nompostamble{%
  \end{multicols}
  \endgroup
}
\renewcommand{\nomgroup}[1]{
\ifthenelse{\equal{G}{#1}}{\item[\textit{Greek symbols}]}{}
\ifthenelse{\equal{S}{#1}}{\item[\textit{Subscripts}]}{}
\ifthenelse{\equal{U}{#1}}{\item[\textit{Superscripts}]}{}}

\setlist{leftmargin=15pt,labelindent=15pt}
\setlist[enumerate]{wide=0pt, leftmargin=15pt, labelwidth=15pt, align=left}

\hypersetup{
  colorlinks   = true, 
  urlcolor     = blue, 
  linkcolor    = blue, 
  citecolor   = blue 
}

\journal{Elsevier}

\begin{document}

\begin{frontmatter}



\title{Versatile two-layer model for thermohydraulic prediction of microchannel heat sinks}


\author[a]{Chih-Hsiang Chen}
\author[a]{Kentaro Yaji\corref{c1}}
\cortext[c1]{Corresponding author}
\ead{yaji@mech.eng.osaka-u.ac.jp}
\affiliation[a]{organization={Department of Mechanical Engineering, Graduate School of Engineering, The University of Osaka},
            addressline={2-1, Yamadaoka}, 
            city={Suita},
	   state={Osaka},
            postcode={ 565-0871}, 
            country={Japan}}

\begin{abstract}
Three-dimensional conjugate heat-transfer simulations provide detailed predictions for microchannel heat sinks but can be computationally demanding when many geometries and operating conditions are evaluated. This study develops a versatile two-layer model (VTLM) that reduces the three-dimensional problem to two coupled two-dimensional fields while retaining the in-plane geometry of the flow passages and fins. Effects associated with the unresolved channel-height direction are incorporated through hydraulic and thermal closures that account for wall friction and the fluid- and fin-side contributions to interlayer heat transfer. The closures are evaluated from the representative channel geometry, flow-development condition, and fin configuration. The VTLM was assessed against three-dimensional reference models across broad ranges of channel aspect ratios and operating conditions, including uniform and nonuniform fin arrangements. Compared with the existing two-layer models, the VTLM showed improved overall agreement with the three-dimensional reference results. The results showed that prediction accuracy was governed mainly by the flow-development condition and channel aspect ratio, from which a conservative screening guideline was proposed for the examined straight-channel configurations. The assessment of finned expansion--contraction channels further supported the applicability of the VTLM to the investigated passages with abrupt width changes. Overall, the VTLM provides a reduced-dimensional framework for thermohydraulic analysis across the range of microchannel heat-sink configurations examined in this study without explicitly resolving the channel-height direction.
\end{abstract}


\end{frontmatter}

\nomenclature{$\mathbf{u}$}{fluid velocity vector ($\rm m/s$)}
\nomenclature{$A$}{area ($\rm m^2$)}
\nomenclature{$\mathrm{AR}$}{channel aspect ratio, \(H/W\)}
\nomenclature{$c_p$}{specific heat capacity ($\rm J/(kg\cdot K)$)}
\nomenclature{$c_{\rm wf}$}{wall-friction coefficient ($\rm Pa\cdot s/m^2$)}
\nomenclature{$D_{\rm h}$}{hydraulic diameter ($\rm m$)}
\nomenclature{$e_{\psi}$}{relative difference in the quantity \(\psi\) (\%)}
\nomenclature{$\mathrm{ER}$}{expansion ratio, \(W/W_{\rm io}\)}
\nomenclature{$f_{\rm D}$}{Darcy friction factor}
\nomenclature{$H$}{common channel and fin height ($\rm m$)}
\nomenclature{$h$}{heat-transfer coefficient ($\rm W/(m^2\cdot K)$)}
\nomenclature{$k$}{thermal conductivity ($\rm W/(m\cdot K)$)}
\nomenclature{$L$}{length ($\rm m$)}
\nomenclature{$L^{+}$}{dimensionless hydraulic flow-development length}
\nomenclature{$N$}{number of fins}
\nomenclature{$\mathrm{Nu}$}{Nusselt number}
\nomenclature{$p$}{pressure ($\rm Pa$)}
\nomenclature{$P$}{perimeter ($\rm m$)}
\nomenclature{$\mathrm{Po}$}{Poiseuille number}
\nomenclature{$\mathrm{Pr}$}{Prandtl number}
\nomenclature{$q''$}{heat flux ($\rm W/m^2$)}
\nomenclature{$Q$}{heat flow ($\rm W$)}
\nomenclature{$\mathrm{Re}$}{Reynolds number}
\nomenclature{$T$}{temperature ($\rm K$)}
\nomenclature{$t$}{representative height or thickness ($\rm m$)}
\nomenclature{$W$}{width ($\rm m$)}
\nomenclature{$x^{*}$}{dimensionless thermal flow-development length}
\nomenclature{$\mathbf{x}$}{in-plane position vector ($\rm m$)}
\nomenclature{$u_c$}{characteristic velocity ($\rm m/s$)}
\nomenclature{$w_{\rm e}$}{entrance-region weighting factor}
\nomenclature{$C_1,C_2$}{coefficients in the Poiseuille-number correlation}
\nomenclature{$C_{t,1},C_{t,2}$}{representative-height closure coefficients}
\nomenclature{$n_{\rm Po}$}{blending exponent in the Poiseuille-number correlation}
\nomenclature{$n_{\rm Nu}$}{blending exponent in the Nusselt-number correlation}
\nomenclature{$n_t$}{blending exponent in the representative-height closure}
\nomenclature{$m$}{fin parameter ($\rm m^{-1}$)}
\nomenclature{$M$}{number of mesh elements}
\nomenclature{$z$}{coordinate measured from the fin base ($\rm m$)}

\nomenclature[G]{$\Lambda$}{dimensionless fin heat-transfer parameter}
\nomenclature[G]{$\zeta_1$}{dimensionless fin-height parameter}
\nomenclature[G]{$\vartheta$}{fin excess temperature ($\rm K$)}
\nomenclature[G]{$\beta$}{rectangular-channel aspect ratio}
\nomenclature[G]{$\lambda_x$}{normalized streamwise fin length}
\nomenclature[G]{$\lambda_y$}{normalized transverse fin width}
\nomenclature[G]{$\mu$}{dynamic viscosity ($\rm Pa\cdot s$)}
\nomenclature[G]{$\rho$}{density ($\rm kg/m^3$)}
\nomenclature[G]{$\tau$}{wall shear stress ($\rm Pa$)}
\nomenclature[G]{$\varphi$}{wall-friction allocation fraction}
\nomenclature[G]{$\Omega$}{computational region}
\nomenclature{$L_{\rm h}$}{streamwise length used for the hydraulic closure ($\rm m$)}
\nomenclature{$R_{\rm th}$}{thermal resistance ($\rm K/W$)}

\nomenclature[S]{$1$}{Layer~1}
\nomenclature[S]{$2$}{Layer~2}
\nomenclature[S]{$\rm bt$}{bottom surface}
\nomenclature[S]{$\rm eff$}{effective}
\nomenclature[S]{$\rm f$}{fluid}
\nomenclature[S]{$\rm fin$}{fin}
\nomenclature[S]{$\rm fp$}{flow passage}
\nomenclature[S]{$\rm GW$}{geometry-weighted}
\nomenclature[S]{$\rm int$}{interlayer}
\nomenclature[S]{$\rm ref$}{reference}
\nomenclature[S]{$\rm s$}{solid}
\nomenclature[S]{$\rm SW$}{shear-weighted}
\nomenclature[S]{$\rm tb$}{top and bottom walls}
\nomenclature[S]{$\rm wf$}{wall friction}
\nomenclature[S]{$\rm b$}{fin base}
\nomenclature[S]{$\rm c$}{characteristic}
\nomenclature[S]{$\rm dev$}{developing flow}
\nomenclature[S]{$\rm e$}{entrance}
\nomenclature[S]{$\rm in$}{inlet}
\nomenclature[S]{$\rm io$}{inlet and outlet}
\nomenclature[S]{$\rm side$}{fin side}
\nomenclature[S]{$\rm top$}{representative Layer~1 height}
\nomenclature[S]{$x$}{streamwise direction}
\nomenclature[S]{$y$}{transverse direction}
\nomenclature[S]{$\rm ax$}{axial}
\nomenclature[S]{$i$}{fin index}

\nomenclature[U]{$\rm 3D$}{three-dimensional reference model}
\nomenclature[U]{$\rm VTLM$}{versatile two-layer model}
\nomenclature[U]{$\rm tot$}{total}

\begin{table*}[!t]
  \begin{framed}
    \printnomenclature
  \end{framed}
\end{table*}

\section{Introduction}

Microchannel heat sinks are widely used for thermal management because their small characteristic dimensions and large heat-transfer area enable efficient heat removal from compact systems ~\cite{zhou2020micro,gao2022fluid,yu2024comprehensive}. Their thermohydraulic behavior is inherently three-dimensional, as coolant flow and thermal development are coupled with heat conduction through the fins and substrate~\cite{fedorov2000three,qu2002analysis,bala2026overview}. Geometric features such as interrupted or segmented fins further modify the local flow and thermal boundary-layer development, increasing the difficulty of accurate performance prediction~\cite{xu2008numerical,naqiuddin2018numerical,ma2025numerical}. Consequently, detailed analysis commonly relies on three-dimensional conjugate heat-transfer models that resolve both the fluid and solid domains.

Three-dimensional conjugate heat-transfer simulations can resolve the coupled fluid and solid fields in detail, but their computational cost becomes substantial when repeated evaluations are required for many geometries or operating conditions~\cite{kulkarni2016multi,alperen2020multi,mohammadbeigi2026numerical}. Physics-based reduced-dimensional modeling provides an alternative for applications requiring repeated thermohydraulic evaluations. In such models, the governing equations are averaged or approximated in the height direction, with additional terms introduced to account for unresolved out-of-plane momentum and heat-transfer effects~\cite {zeng2018experimental,zou2022topology,xie2022topology}. These models retain conservation equations for the flow and temperature fields in a lower-dimensional physical domain while substantially reducing the computational cost.

Porous-medium and averaging approaches have been developed primarily for microchannel heat sinks composed of regular channel-fin arrays. In porous-medium formulations, the repeated channel-fin structure is homogenized as an equivalent porous domain, whereas averaging approaches describe its thermohydraulic behavior using spatially averaged momentum and energy equations~\cite{zhao2002analysis,kim2006averaging}. A subsequent formulation incorporated entrance effects by distinguishing between developed and developing-flow regimes and was assessed against a three-dimensional conjugate heat-transfer model~\cite{kewalramani2017modeling}. These approaches provide efficient predictions by representing regularly arranged channels and fins through averaged geometric and thermohydraulic properties.

Reduced-dimensional models have also been developed to retain more complex in-plane channel and fin arrangements. Harpole and Eninger formulated an early two-dimensional flow and thermal model for a manifold microchannel heat exchanger, in which unresolved viscous resistance and fluid-solid heat exchange were represented through effective relations~\cite{harpole1991micro}. Mandel et al. subsequently developed a 2.5-D model for manifold microchannels that retained the inertial terms and incorporated developing-flow effects within a two-dimensional computational representation. The model substantially reduced the computational cost relative to fully three-dimensional simulations, although its accuracy deteriorated when secondary three-dimensional flow structures became significant~\cite{mandel2018modeling}. Kamal and Dewan adopted a two-dimensional model for high-aspect-ratio interrupted microchannels after comparing its predictions with three-dimensional simulations~\cite{kamal2017analysis}. In this case, the validity of the reduced representation relied on sufficiently high aspect ratios to limit variations in the unresolved direction. These studies demonstrate that planar computational models can accommodate relatively complex in-plane flow passages, although their applicability depends on the significance of the unresolved three-dimensional effects.

\begin{figure*}[htbp]
\centering
\includegraphics[width=0.8\textwidth]{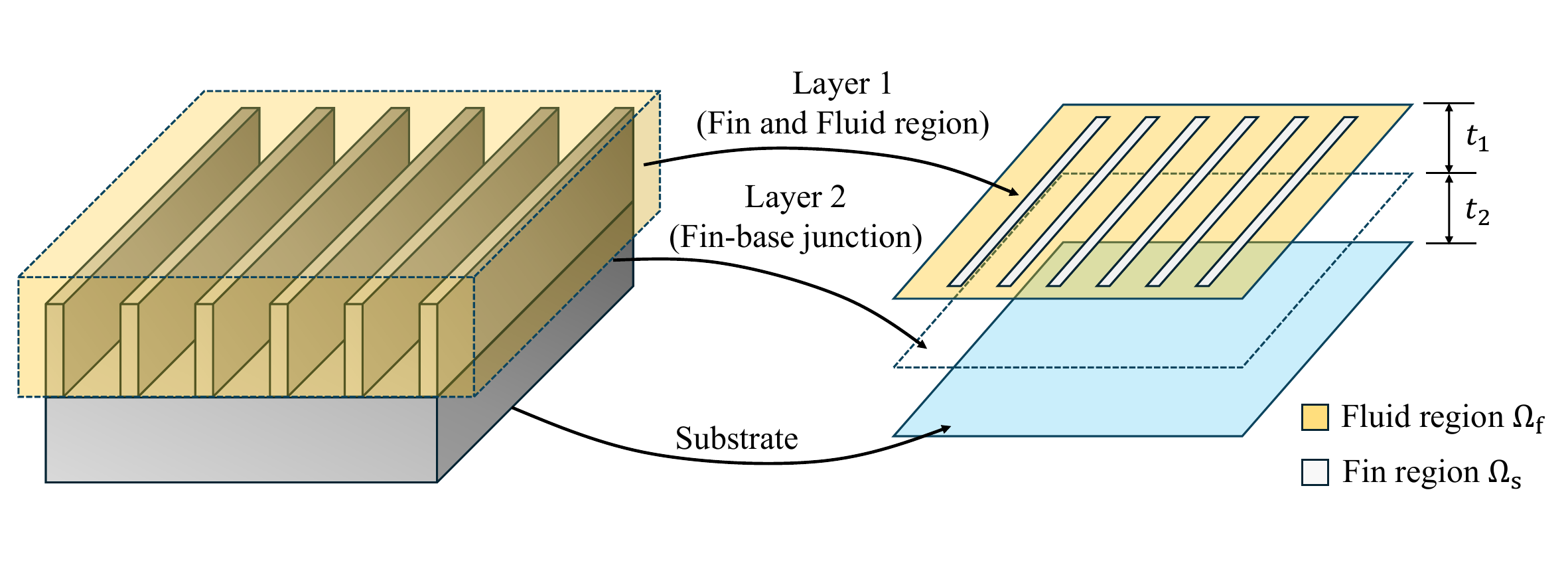} 
\caption{Schematic representation of the VTLM. The three-dimensional microchannel heat sink is reduced to two coupled planar fields: Layer~1, which contains the fluid and fin regions (\(\Omega_{\rm f}\) and \(\Omega_{\rm s}\)), and Layer~2, which represents the fin-base junction and substrate region. The representative height associated with Layer~1 and the substrate thickness are denoted by \(t_1\) and \(t_2\), respectively.}
\label{Fig.1}
\end{figure*}

The need for computationally efficient reduced-dimensional models is particularly pronounced in topology optimization (TO), where the governing equations must be solved repeatedly as the fluid--solid distribution evolves. TO provides a systematic approach for generating flow-channel layouts by optimizing the spatial distribution of fluid and solid regions within a prescribed design domain~\cite{bendsoe2004topology}. Following the density-based formulation introduced by Borrvall and Petersson for Stokes flow~\cite{borrvall2003topology}, fluid topology optimization was extended to more general flow problems~\cite{olesen2006high,deng2011topology,dilgen2018topology} and coupled thermal--fluid systems~\cite{dede2009multiphysics,yaji2015topology,chen2025topology}. To reduce the computational cost associated with these repeated analyses, pseudo-three-dimensional and two-layer models have also been introduced for heat-sink design and topology optimization. In these models, the three-dimensional conjugate heat-transfer problem is reduced to coupled planar domains, while effects associated with the unresolved direction are represented through simplifying assumptions or effective relations. Haertel et al. developed a pseudo-three-dimensional model comprising two thermally coupled planar layers: a conducting base layer and a thermofluid design layer~\cite{haertel2018topology}. Comparison with a three-dimensional model showed that the omission of top- and bottom-wall friction introduced systematic differences in the predicted flow rate, although the thermal-resistance predictions remained in good agreement. A related formulation was adopted by Zeng and Lee for heat-sink topology optimization, and the resulting designs were subsequently assessed using three-dimensional simulations and experiments~\cite{zeng2019topology}. Yan et al. proposed a two-layer model for the topology optimization of microchannel heat sinks based on a plane-flow assumption, in which coupled fluid and solid thermal fields were used to approximate three-dimensional conjugate heat transfer~\cite{yan2019topology}. Zhao et al.~\cite{zhao2021topology} introduced a multilayer thermofluid model in which adaptive temperature profiles in the height direction were employed to better represent out-of-plane heat-transfer effects. A related multilayer formulation was later used for heat-sink topology optimization under turbulent forced-convection conditions~\cite{zhao2021topology2}. More recently, Jia et al. incorporated a multilayer representation into a topology optimization framework for microfluidic heat sinks with non-Newtonian fluids~\cite{jia2026multilayer}.

A useful distinction among the reduced-dimensional modeling approaches reviewed above is how the in-plane channel-fin geometry is represented. Porous-medium and averaging approaches replace the detailed channel-fin structure with an equivalent averaged representation and have been developed mainly for regularly arranged channel-fin configurations~\cite{zhao2002analysis,kim2006averaging,kewalramani2017modeling}. By contrast, several reduced-dimensional formulations retain the in-plane flow-passage geometry within the computational domain, including two-layer models used in heat-sink design and topology optimization~\cite{haertel2018topology,zeng2019topology,yan2019topology}. However, their applicability tends to depend on the channel-height regime. Models based on a plane-flow assumption are most appropriate for low-aspect-ratio channels, where variations across the channel height remain limited~\cite{yan2019topology,zhao2021topology,zhao2021topology2,jia2026multilayer}. In contrast, the limitation associated with neglecting top- and bottom-wall friction becomes less significant at high aspect ratios, where the relative contribution of these walls to the total hydraulic resistance is reduced~\cite{haertel2018topology,zeng2019topology}. Simplified thermal treatments may likewise lose accuracy as the channel geometry and thermal-development condition vary~\cite{zeng2019topology}.

The applicability of existing reduced-dimensional formulations is generally limited to particular aspect-ratio regimes. A formulation capable of retaining the in-plane geometry while accommodating a wide range of aspect ratios remains needed, and its applicable range should be clearly established. To address this need, the present study develops a versatile two-layer model (VTLM) for thermohydraulic prediction of microchannel heat sinks over a broad range of channel aspect ratios. The original three-dimensional conjugate heat-transfer problem is represented by two coupled two-dimensional fields: one describing the in-plane fluid and fin regions and the other accounting for heat conduction near the fin-base junction and within the substrate. The VTLM accounts for the unresolved height-direction effects through three closure models evaluated according to the representative geometry and development conditions. The wall-friction closure accounts for momentum loss at the unresolved top and bottom walls, the fluid-side heat-transfer closure incorporates flow-passage geometry and thermal development, and the fin-side effective heat-transfer closure represents both axial fin conduction and lateral heat transfer to the coolant. The prediction accuracy and applicable range of the VTLM are systematically assessed against three-dimensional reference simulations over a wide range of channel aspect ratios and fin configurations. In addition, the models of Yan et al.~\cite{yan2019topology} and Zeng and Lee~\cite{zeng2019topology} are compared at representative aspect ratios to evaluate the VTLM relative to existing two-layer formulations. Although both models were developed in the context of topology optimization, the present comparison concerns only their underlying thermohydraulic analysis formulations.

The remainder of this paper is organized as follows. Section~\ref{Mathematical model} presents the assumptions, governing equations, and thermohydraulic closure models of the VTLM. Section~\ref{Numerical implementation} describes the numerical implementation of the VTLM and the three-dimensional simulations. Section~\ref{Results and discussion} presents the numerical assessment for different fin-array configurations and flow-passage geometries and discusses the applicability of the proposed model. Finally, Section~\ref{Conclusion} summarizes the conclusions.

\section{Mathematical model}
\label{Mathematical model}

\subsection{Assumptions}
\label{Assumptions}

As illustrated in Fig.~\ref{Fig.1}, the VTLM reduces the three-dimensional conjugate heat-transfer domain to two coupled planar fields. Layer~1 resolves the in-plane coolant and fin regions, whereas Layer~2 represents heat conduction near the fin--base junction and within the substrate. Variations in the channel-height direction are not resolved explicitly. Instead, their effects on momentum loss and heat exchange between the two layers are incorporated through the closure models described in Section~\ref{sec:closure-models}. The formulation is developed under the following conditions:

\begin{itemize}
\item The coolant flow is steady, laminar, incompressible, and Newtonian. The thermophysical properties of the coolant and solid materials are assumed to be constant.
\item The in-plane channel-fin geometry is extruded uniformly in the height direction. This assumption excludes configurations with spatially varying channel or fin heights, such as tapered or stepped profiles.
\item The configurations have a predominant mean flow direction, with the inlet and outlet located on opposite sides of the computational domain. This direction is used to define the representative streamwise dimensions and flow-development lengths required by the closure models.
\item Heat conduction along the fin height is represented by a one-dimensional fin model with an adiabatic tip~\cite{kim2010analytical}.
\end{itemize}

\subsection{Governing equations}
\label{Governing equations}

The VTLM is formulated on the two-dimensional plane shown in Fig.~\ref{Fig.1}. Layer~1 is divided into the fluid region \(\Omega_{\rm f}\) and the fin region \(\Omega_{\rm s}\), whereas Layer~2 spans the entire in-plane footprint and represents the fin-base junction and the underlying substrate. The velocity and pressure fields are solved only in $\Omega_{\rm f}$. The temperatures in the fluid and fin regions of Layer~1 are denoted by $T_{1,\rm f}$ and $T_{1,\rm s}$, respectively, while $T_2$ denotes the temperature in Layer~2. The common channel and fin height is denoted by \(H\), while \(t_1\) and \(t_2\) denote the representative height associated with Layer~1 and the substrate thickness, respectively. The former also specifies the representative height of Layer~1 in the fin-side closure, whereas the latter corresponds to the substrate thickness. All spatial differential operators appearing below act in the in-plane directions. For steady incompressible flow, mass conservation in the fluid region is expressed as
\begin{equation}
\nabla \cdot \mathbf{u}=0
\qquad \text{in } \Omega_{\rm f},
\label{eq:continuity}
\end{equation}
where $\mathbf{u}$ is the fluid velocity. The corresponding momentum equation is
\begin{equation}
\rho_{\rm f}
\left(
\mathbf{u}\cdot\nabla
\right)\mathbf{u}
=
-\nabla p
+
\nabla\cdot
\left[
\mu_{\rm f}
\left(
\nabla\mathbf{u}
+
\nabla\mathbf{u}^{\mathsf T}
\right)
\right]
-
c_{\rm wf}\mathbf{u}
\qquad \text{in } \Omega_{\rm f},
\label{eq:momentum}
\end{equation}
where $\rho_{\rm f}$ is the fluid density, $\mu_{\rm f}$ is the dynamic viscosity, and $p$ is the pressure. The final term represents the momentum loss associated with viscous shear on the unresolved top and bottom channel walls. The wall-friction coefficient $c_{\rm wf}$ is evaluated using the closure model presented in Section~\ref{wall-friction-coef}. The energy equation in the fluid region of Layer~1 is given by
\begin{equation}
\rho_{\rm f} c_{p,\rm f}
\mathbf{u}\cdot\nabla T_{1,\rm f}
=
\nabla\cdot
\left(
k_{\rm f}\nabla T_{1,\rm f}
\right)
+
\frac{h_{\rm f}}{H}
\left(
T_2-T_{1,\rm f}
\right)
\qquad \text{in } \Omega_{\rm f},
\label{eq:energy-layer1-fluid}
\end{equation}
where $c_{p,\rm f}$ and $k_{\rm f}$ are the specific heat and thermal conductivity of the fluid. The coefficient $h_{\rm f}$ represents the effective heat transfer between the fluid region of Layer~1 and Layer~2. Within the fin region, heat transport in Layer~1 is governed by
\begin{equation}
\nabla\cdot
\left(
k_{\rm s}\nabla T_{1,\rm s}
\right)
+
\frac{h_{\rm s}}{H}
\left(
T_2-T_{1,\rm s}
\right)
=
0
\qquad \text{in } \Omega_{\rm s},
\label{eq:energy-layer1-solid}
\end{equation}
where $k_{\rm s}$ is the thermal conductivity of the solid material and $h_{\rm s}$ is the fin-side effective heat-transfer coefficient. The latter accounts for heat conduction along the fin height and heat transfer from the fin side surfaces, as described in Section~\ref{subsec:fin_side_htc}. To express the coupling with Layer~2 in a unified form, the Layer~1 temperature and the interlayer coefficient are defined as
\begin{equation}
\left(
T_1,h_{\rm int}
\right)
=
\begin{cases}
\left(
T_{1,\rm f},h_{\rm f}
\right),
& \mathbf{x}\in\Omega_{\rm f},
\\[4pt]
\left(
T_{1,\rm s},h_{\rm s}
\right),
& \mathbf{x}\in\Omega_{\rm s},
\end{cases}
\label{eq:interlayer-piecewise}
\end{equation}
where $\mathbf{x}$ denotes the in-plane position vector. The energy balance in Layer~2 is then written as
\begin{equation}
\nabla\cdot
\left(
k_{\rm s}\nabla T_2
\right)
+
\frac{q''_{\rm bt}}{t_2}
-
\frac{h_{\rm int}}{t_2}
\left(
T_2-T_1
\right)
=
0,
\label{eq:energy-layer2}
\end{equation}
where $q''_{\rm bt}$ is the heat flux supplied at the bottom surface. The second term represents the applied heat input distributed over the thickness $t_2$, whereas the third term describes heat exchange between the two planar fields. When the bottom-surface temperature $T_{\rm bt}$ is required, the in-plane temperature variation is already represented by the Layer~2 temperature field. The unresolved temperature drop across the substrate thickness is reconstructed locally using a one-dimensional conduction relation, following Zeng and Lee~\cite{zeng2019topology}:
\begin{equation}
T_{\rm bt}
=
T_2
+
\frac{q''_{\rm bt}t_2}{k_{\rm s}}.
\label{eq:bottom-temperature}
\end{equation}

\subsection{Thermohydraulic closure models}
\label{sec:closure-models}

\subsubsection{Wall-friction closure}
\label{wall-friction-coef}

The in-plane momentum equation in Layer~1 does not resolve the velocity gradients normal to the top and bottom channel walls. The associated viscous resistance is represented by the linear momentum sink \(-c_{\rm wf}\mathbf{u}\). By matching this sink with the pressure gradient of laminar flow through a rectangular duct, the wall-friction coefficient is expressed as
\begin{equation}
c_{\rm wf}
=
\frac{\varphi_{\rm tb}\,\mathrm{Po}\,\mu_{\rm f}}
     {2D_{\rm h}^{2}},
\label{eq:sfc_coefficient}
\end{equation}
where \(\mathrm{Po}\) is the Poiseuille number, and \(\varphi_{\rm tb}\) denotes the fraction of the total wall friction attributed to the unresolved top and bottom walls. For a rectangular channel of height \(H\) and width \(W\), hydraulic diameter \(D_{\rm h}\) can be expressed as follows:
\begin{equation}
D_{\rm h}
=
\frac{2HW}{H+W}.
\label{eq:sfc_hydraulic_diameter}
\end{equation}

To account for both hydrodynamically developing and fully developed flow, the Poiseuille number is evaluated using the composite correlation proposed by Muzychka and Yovanovich~\cite{muzychka2002laminar}, expressed in the Churchill--Usagi form~\cite{churchill1972general}:
\begin{equation}
\mathrm{Po}
=
f_{\rm D}\mathrm{Re}
=
\left[
C_1^{\,n_{\rm Po}}
+
\left(
\frac{C_2}{\sqrt{L^{+}}}
\right)^{n_{\rm Po}}
\right]^{1/n_{\rm Po}},
\label{eq:sfc_po}
\end{equation}
where \(f_{\rm D}\) is the Darcy friction factor, \(n_{\rm Po}=2\), 
\(C_2=13.76\), and
\begin{equation}
L^{+}
=
\frac{L_{\rm h}}{D_{\rm h}\mathrm{Re}},
\qquad
\mathrm{Re}
=
\frac{\rho_{\rm f} u_c D_{\rm h}}{\mu_{\rm f}},
\label{eq:sfc_dimensionless_length}
\end{equation}
where \(L_{\rm h}\) denotes the streamwise length of the flow-passage segment over which the hydraulic closure is evaluated, and \(u_c\) is the characteristic velocity, defined as the mean velocity at the inlet. The fully developed contribution \(C_1\) is obtained from the rectangular-duct relation of Shah and London~\cite{shah2014laminar}:
\begin{equation}
\begin{split}
C_1
=
96\bigl(
&1
-1.3553\beta
+1.9467\beta^{2}
-1.7012\beta^{3} \\
&+0.9564\beta^{4}
-0.2537\beta^{5}
\bigr),
\end{split}
\label{eq:sfc_C1}
\end{equation}
where \(\beta\) denotes the cross-sectional aspect ratio, defined as
\begin{equation}
\beta
=
\frac{\min(H,W)}{\max(H,W)},
\qquad
0<\beta\leq 1.
\label{eq:sfc_beta}
\end{equation}

Consequently, Eq.~\eqref{eq:sfc_po} adjusts the predicted friction according to both the duct geometry and hydrodynamic development. The fraction \(\varphi_{\rm tb}\) is evaluated by blending the geometry-weighted estimate \(\varphi_{\rm GW}\) for the entrance region with the shear-weighted estimate \(\varphi_{\rm SW}\) for fully developed flow as follows:
\begin{equation}
\varphi_{\rm tb}
=
w_{\rm e}\varphi_{\rm GW}
+
\left(1-w_{\rm e}\right)\varphi_{\rm SW},
\qquad
w_{\rm e}
=
\min\left(\frac{L_{\rm e}}{L_{\rm h}},1\right),
\label{eq:sfc_partition}
\end{equation}
where \(w_{\rm e}\) is the entrance-region weighting factor and
\(L_{\rm e}=0.05\,\mathrm{Re}\,D_{\rm h}\) is the estimated hydrodynamic entrance length~\cite{lee2005investigation}. Near the inlet, the wall-shear profile has not yet approached its fully developed distribution. Therefore, the top- and bottom-wall contribution is approximated from the corresponding fraction of the wetted perimeter:
\begin{equation}
\varphi_{\rm GW}
=
\frac{2W}{2(H+W)}
=
\frac{W}{H+W}.
\label{eq:sfc_phi_GW}
\end{equation}

Farther downstream, the allocation between the two pairs of opposing walls is instead determined from their mean shear stresses. Let \(\bar{\tau}_{H}\) and \(\bar{\tau}_{W}\) denote the magnitudes of the mean shear stresses on the wall groups whose individual lengths are \(H\) and \(W\), respectively. Since the top and bottom surfaces belong to the \(W\)-wall group, their fully developed contribution is written as
\begin{equation}
\varphi_{\rm SW}
=
\frac{W\bar{\tau}_{W}}
     {H\bar{\tau}_{H}+W\bar{\tau}_{W}}.
\label{eq:sfc_phi_SW}
\end{equation}

Following the rectangular-duct solution reported by Delplace~\cite{delplace2018laminar}, the mean shear-stress magnitudes are evaluated from the following expressions:
\begin{equation}
\bar{\tau}_{H}
=
\frac{16\Delta p\,H}{\pi^{3}L_{\rm h}}
\sum_{j=1,3,\ldots}^{\infty}
\frac{1}{j^{3}}
\tanh\left(
\frac{j\pi W}{2H}
\right),
\label{eq:sfc_tau_H}
\end{equation}
and
\begin{equation}
\bar{\tau}_{W}
=
\frac{16\Delta p\,W}{\pi^{3}L_{\rm h}}
\sum_{j=1,3,\ldots}^{\infty}
\frac{1}{j^{3}}
\tanh\left(
\frac{j\pi H}{2W}
\right),
\label{eq:sfc_tau_W}
\end{equation}
where \(\Delta p\) is the pressure drop over the streamwise length \(L_{\rm h}\), and the summations are taken over the positive odd integers \(j\). 

\begin{figure}[h]
    \centering
    \includegraphics[width=1.0\linewidth]{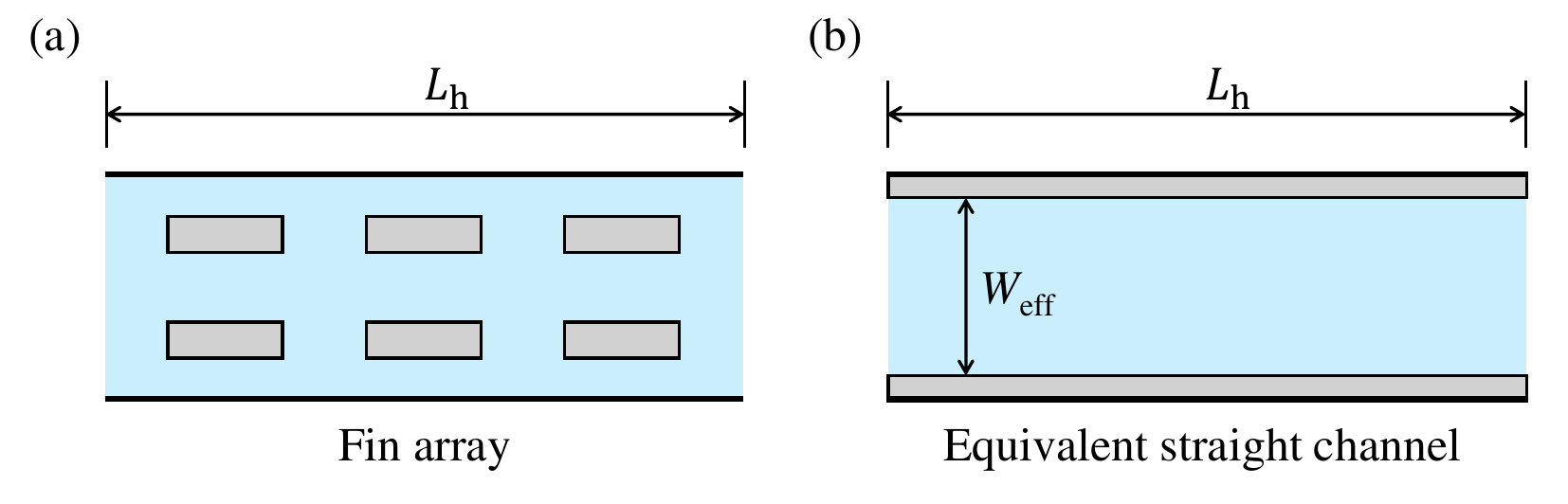}
\caption{Equivalent straight-channel representation used in the wall-friction closure for a fin-array passage: (a) original fin-array passage and (b) equivalent straight channel.}
    \label{fig:sfc_equivalent_channel}
\end{figure}

For channels containing fin arrays, the flow passage cannot be uniquely assigned a rectangular cross-section. Although the hydraulic resistance associated with the fin boundaries is resolved directly in Layer~1, the closure still requires a representative rectangular cross-section to evaluate the top- and bottom-wall contribution. Accordingly, the flow passage is mapped to an area-preserving straight channel, as shown in Fig.~\ref{fig:sfc_equivalent_channel}. Its effective width is evaluated as $W_{\rm eff}=A_{\rm fp}/L_{\rm h}$ where \(A_{\rm fp}\) is the flow-passage area of the corresponding segment. The associated aspect ratio and hydraulic diameter are
\begin{equation}
\beta_{\rm eff}
=
\frac{\min(H,W_{\rm eff})}
     {\max(H,W_{\rm eff})},
\qquad
D_{{\rm h},\rm eff}
=
\frac{2HW_{\rm eff}}
     {H+W_{\rm eff}}.
\label{eq:sfc_effective_geometry}
\end{equation}

All quantities in the wall-friction closure are then evaluated for the equivalent rectangular channel by replacing \(W\), \(\beta\), and \(D_{\rm h}\) with \(W_{\rm eff}\), \(\beta_{\rm eff}\), and \(D_{{\rm h},\rm eff}\), respectively. Accordingly, \(\mathrm{Re}\), \(L^{+}\), and \(L_{\rm e}\) are evaluated using \(D_{{\rm h},\rm eff}\). This equivalent representation provides the global geometric quantities required by the straight-channel friction closure.

\subsubsection{Fluid-side heat transfer coefficient}
\label{subsec:fluid_htc}

The fluid-side contribution to the interlayer heat exchange is evaluated by distinguishing the portions associated with the fin sidewalls and the flow-passage surface. These two portions generally have different streamwise thermal-development lengths and are treated separately before being combined into a single coefficient. The area-averaged fluid-side heat transfer coefficient is defined as
\begin{equation}
h_{\rm f}
=
\frac{
h_{\rm f,fin} A_{\rm fin}^{\rm tot}
+
h_{\rm f,fp} A_{\rm fp}
}{
A_{\rm fin}^{\rm tot}
+
A_{\rm fp}
},
\label{eq:hf_effective}
\end{equation}
where \(h_{\rm f,fin}\) and \(h_{\rm f,fp}\) denote the heat-transfer coefficients assigned to the fin and flow-passage contributions, respectively. The quantity \(A_{\rm fin}^{\rm tot}\) denotes the total in-plane fin area. Both coefficients are evaluated using a laminar developing-flow correlation for a rectangular channel, with separate thermal-development lengths for the fin-sidewall and flow-passage contributions:
\begin{equation}
h_{\rm f,fin}
=
\frac{
\mathrm{Nu}_{\rm dev}(x_{\rm fin}^{*})\,k_{\rm f}
}{
D_{\rm h}
},
\qquad
h_{\rm f,fp}
=
\frac{
\mathrm{Nu}_{\rm dev}(x_{\rm fp}^{*})\,k_{\rm f}
}{
D_{\rm h}
}.
\label{eq:hf_components}
\end{equation}

The corresponding dimensionless thermal-development lengths are shown below:
\begin{equation}
x_{\rm fin}^{*}
=
\frac{
L_{\rm fin}
}{
\mathrm{Re}\,\mathrm{Pr}\,D_{\rm h}
},
\qquad
x_{\rm fp}^{*}
=
\frac{
L_{\rm fp}
}{
\mathrm{Re}\,\mathrm{Pr}\,D_{\rm h}
},
\label{eq:thermal_development_lengths}
\end{equation}
where \(\mathrm{Pr}=\mu_{\rm f}c_{p,\rm f}/k_{\rm f}\) is the Prandtl number, while \(L_{\rm fin}\) and \(L_{\rm fp}\) denote the representative streamwise fin length and the streamwise flow-passage length, respectively. The developing-flow Nusselt number is calculated using the correlation proposed by Muzychka and Yovanovich~\cite{muzychka2002laminar2}:
\begin{equation}
\mathrm{Nu}_{\rm dev}(x^{*})
=
\left\{
\left[
0.4886
\left(
\frac{C_1}{x^{*}}
\right)^{1/3}
\right]^{n_{\rm Nu}}
+
\left[
0.1144
\frac{C_1}
{\sqrt{\pi\beta^{0.1}}}
\right]^{n_{\rm Nu}}
\right\}^{1/n_{\rm Nu}},
\label{eq:nu_developing}
\end{equation}
where \(C_1\) is the fully developed rectangular-duct coefficient defined in Eq.~\eqref{eq:sfc_C1}, \(\beta\) is the channel aspect ratio defined in Eq.~\eqref{eq:sfc_beta}, and \(n_{\rm Nu}=5\) is the blending exponent. The correlation was originally formulated for a rectangular channel heated on all four walls. In the present model, it is applied as an approximation to the three-sided heating configuration.

\begin{figure}[h]
    \centering
    \includegraphics[width=1.0\linewidth]{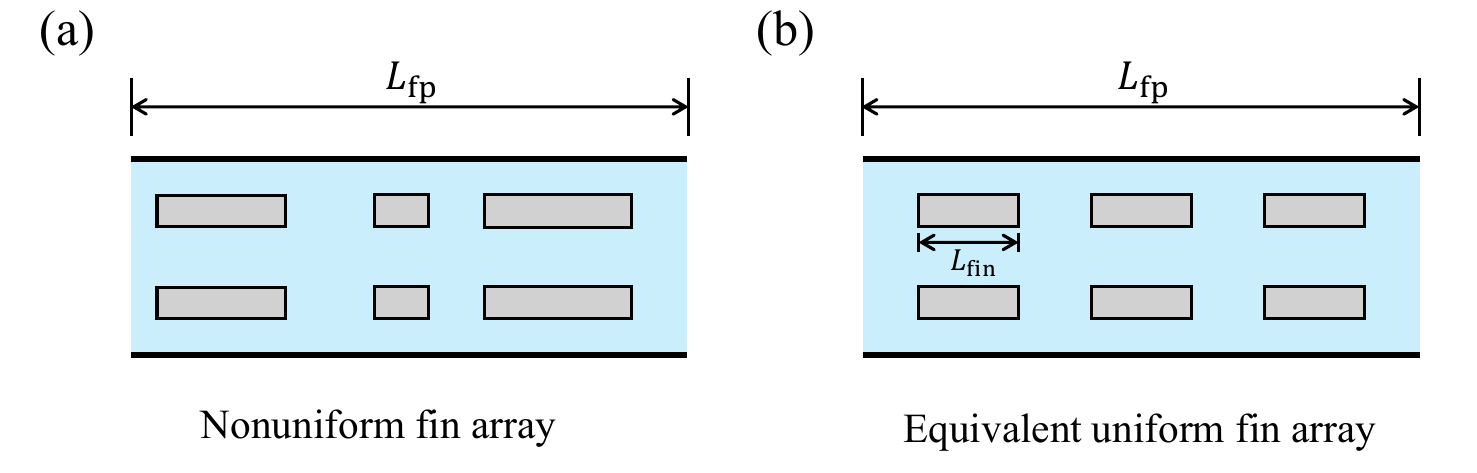}
\caption{Equivalent uniform fin-array representation for evaluating the fin-side thermal-development length: (a) nonuniform fin array and (b) equivalent uniform fin array characterized by \(L_{\rm fin}\).}
    \label{fig:equivalent_fin_array}
\end{figure}

For a uniform fin array, \(L_{\rm fin}\) is the streamwise fin length. For a nonuniform fin array, the geometry is mapped to an equivalent uniform fin array, as illustrated in Fig.~\ref{fig:equivalent_fin_array}, and \(L_{\rm fin}\) is evaluated as follows:
\begin{equation}
L_{\rm fin}
=
\frac{
2A_{\rm fin}^{\rm tot}
}{
P_{\rm fin}^{\rm tot}
-
2L_{\rm fin}^{\rm tot}
},
\label{eq:equivalent_fin_length}
\end{equation}
where \(P_{\rm fin}^{\rm tot}\) and \(L_{\rm fin}^{\rm tot}\) denote the total in-plane fin perimeter and the sum of the streamwise fin lengths, respectively.

\subsubsection{Fin-side effective heat transfer coefficient}
\label{subsec:fin_side_htc}

The fluid-side coefficient introduced in Section~\ref{subsec:fluid_htc} describes heat transfer from the fin sidewalls to the coolant but does not account for heat conduction within
the fin. A fin-side effective heat transfer coefficient \(h_{\rm s}\) is introduced to represent the combined influence of axial fin conduction and lateral heat transfer to the fluid.

\begin{figure}[h]
    \centering
    \includegraphics[width=0.8\linewidth] {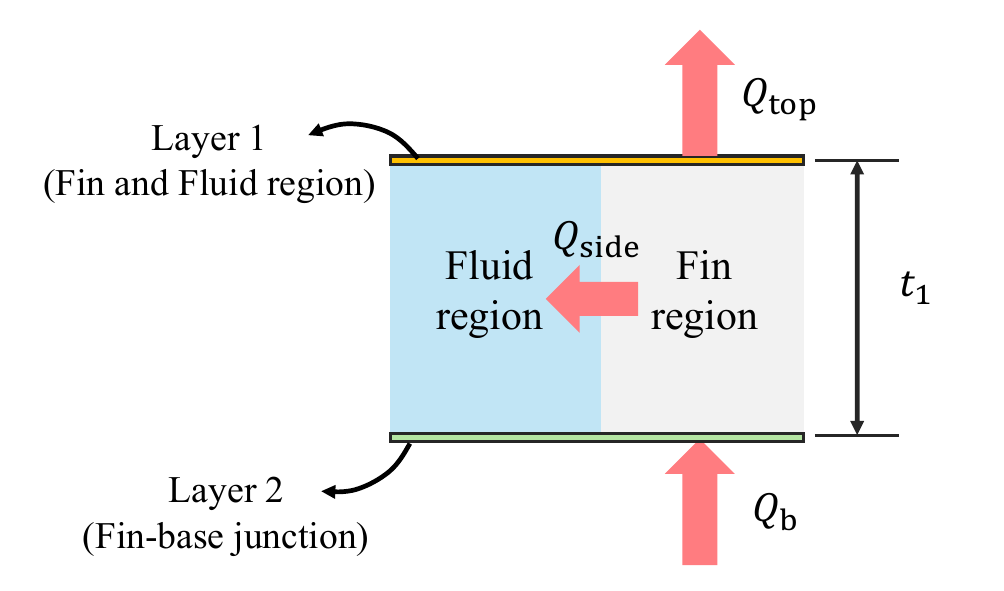}
    \caption{Schematic of the fin-side thermal closure. Heat entering the fin from Layer~2 is divided into lateral heat transfer to the coolant and the remaining axial heat flow evaluated at the representative height \(t_1\).}
    \label{fig:fin_side_closure}
\end{figure}

Fig.~\ref{fig:fin_side_closure} illustrates the corresponding heat-flow representation. Heat entering the fin from Layer~2 is denoted by \(Q_{\rm b}\). A portion of this heat is transferred laterally to the fluid as \(Q_{\rm side}\), while the remaining axial heat flow \(Q_{\rm top}\) reaches the representative Layer~1 height \(t_1\). The fin-side effective heat transfer coefficient is defined from this remaining heat flow as

\begin{equation}
h_{\rm s}
=
\frac{
Q_{\rm top}
}{
A_{\rm fin}^{\rm tot}
\left[
T_2-T_{\rm fin}(t_1)
\right]
},
\label{eq:hs_definition}
\end{equation}
where \(T_2=T_{\rm fin}(0)\) is the temperature at the fin-base junction, and \(T_{\rm fin}(t_1)\) is the fin temperature at the representative height. Under the uniform-extrusion assumption, \(A_{\rm fin}^{\rm tot}\) also serves as the total cross-sectional area normal to heat conduction along the fin height. Here, \(t_1\) denotes the representative height of Layer~1 and is an effective closure parameter rather than a physical geometric dimension.

Using a one-dimensional fin approximation with constant properties and an adiabatic fin tip, Eq.~\eqref{eq:hs_definition} can be written in closed form as
\begin{equation}
h_{\rm s}
=
\frac{k_{\rm s}}{H}
\frac{
\Lambda \sinh \zeta_1
}{
\cosh \Lambda-\cosh \zeta_1
},
\label{eq:hs_closed_form}
\end{equation}
where
\begin{equation}
\Lambda
=
H
\sqrt{
\frac{
h_{\rm f,fin}P_{\rm fin}^{\rm tot}
}{
k_{\rm s}A_{\rm fin}^{\rm tot}
}
},
\qquad
\zeta_1
=
\Lambda
\left(
1-\frac{t_1}{H}
\right).
\label{eq:fin_side_parameters}
\end{equation}
Here, \(H\) denotes the common channel and fin height. The coefficient \(h_{\rm f,fin}\) is evaluated using the fluid-side closure described in Section~\ref{subsec:fluid_htc}. The detailed derivation of Eqs.~\eqref{eq:hs_closed_form} and \eqref{eq:fin_side_parameters} is provided in~\ref{app:fin_side_htc_derivation}.

\begin{figure}[h]
    \centering
    \includegraphics[width=1.0\linewidth]{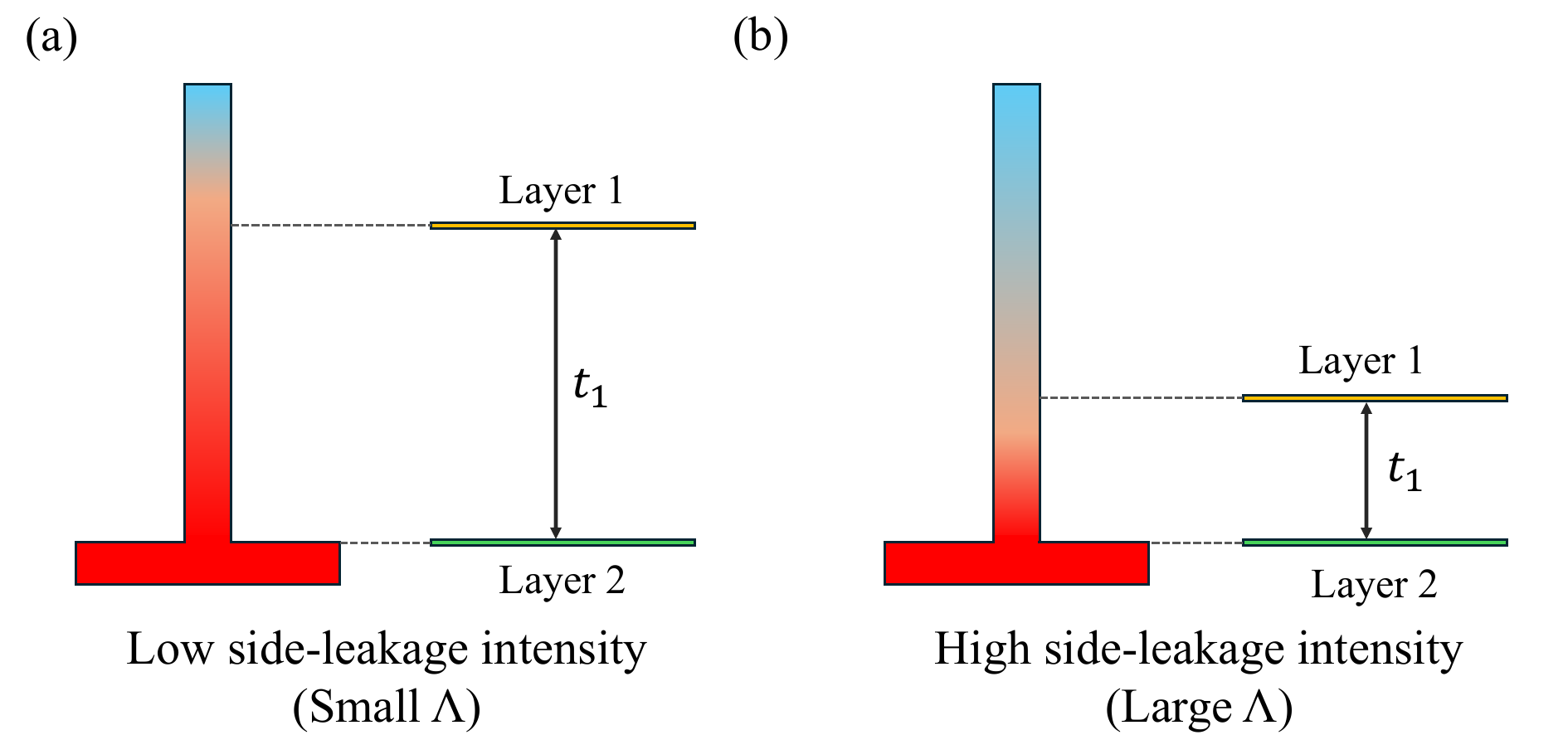}
\caption{Effect of the lateral heat-transfer intensity on the representative Layer~1 height: (a) weak lateral heat transfer with small \(\Lambda\) and (b) strong lateral heat transfer with large \(\Lambda\). Increasing \(\Lambda\) shifts the representative Layer~1 height toward the fin base.}
    \label{fig:representative_height}
\end{figure}

The dimensionless parameter \(\Lambda\) characterizes lateral heat transfer relative to axial conduction within the fin. Because the effective location of the interlayer heat exchange changes with \(\Lambda\), the representative height \(t_1\) is not prescribed as a fixed fraction of \(H\). As illustrated in Fig.~\ref{fig:representative_height}, weak lateral heat transfer allows heat to travel farther toward the fin tip and gives a larger \(t_1\). By contrast, strong lateral heat transfer concentrates the effective exchange closer to the fin base and leads to a smaller \(t_1\). The representative-height ratio is evaluated using the following Churchill--Usagi-type blending relation:
\begin{equation}
\frac{t_1}{H}
=
\left[
\left(
C_{t,1}
\right)^{-n_t}
+
\left(
\frac{C_{t,2}}{\Lambda}
\right)^{-n_t}
\right]^{-1/n_t},
\label{eq:representative_height}
\end{equation}
where closure parameters are set to \(n_t\approx1.4\), \(C_{t,1}=0.8\), and \(C_{t,2}=0.693\). These values were calibrated using the baseline straight-channel dataset and were retained without further adjustment in the subsequent cases. For configurations without fins, the solid region \(\Omega_{\rm s}\) is absent from Layer~1. Accordingly, the fin-side coefficient \(h_{\rm s}\) and the representative-height relation in Eq.~\eqref{eq:representative_height} are not used. Interlayer heat transfer is represented solely by the fluid-side coefficient \(h_{\rm f}\).

\section{Numerical implementation}
\label{Numerical implementation}

The VTLM and the corresponding three-dimensional reference model were implemented using the finite-element method in COMSOL Multiphysics 6.1. In both models, the velocity and pressure fields were solved using the Laminar Flow interface, while conjugate heat transfer in the fluid and solid regions was evaluated using the Heat Transfer in Solids and Fluids interface. For the VTLM, Layer~1 and Layer~2 were constructed as separate two-dimensional components, with Layer~2 modeled using the Heat Transfer in Solids interface. The velocity and pressure fields were discretized using equal-order first-order finite elements (P1+P1), while linear finite elements were used for the temperature field. Consistent streamline-diffusion and crosswind-diffusion stabilization were applied to both the flow and heat-transfer equations. The wall-friction sink term was introduced into the Layer~1 momentum equation, and the interlayer heat-transfer terms were applied with opposite signs in the energy equations of the two layers, using \(h_{\rm f}\) and \(h_{\rm s}\) in the flow-passage and fin regions, respectively. The three-dimensional reference model was resolved directly without reduced-dimensional hydraulic or thermal closure terms. All calculations were performed under stationary conditions. The same material properties and boundary conditions were imposed in the VTLM and three-dimensional calculations for each case. A relative solver tolerance of \(1.0\times10^{-4}\) was used throughout the study. All simulations were performed on the same workstation equipped with a 13th Gen Intel Core i7-13700 processor. The global energy balance was also verified by comparing the imposed heat input with the coolant enthalpy increase, \(\dot{m}c_{p,\rm f}(T_{\rm out}-T_{\rm in})\), confirming that the energy imbalance was negligible for the conjugate heat-transfer cases.

\section{Results and discussion}
\label{Results and discussion}

\subsection{Assessment of the three-dimensional reference model}
\label{subsec:reference_model_verification}

Before its use as the reference model for evaluating the VTLM, the three-dimensional numerical implementation was examined through a mesh-independence study and comparison with analytical solutions. Both tests employed a straight square channel containing only the fluid domain. No solid domain and conjugate heat transfer were included.

The channel had a width and height of \(W=H=3.0~\mathrm{mm}\) and a length of \(L=45.0~\mathrm{mm}\). For this benchmark, the streamwise development length in \(L^{+}\) was taken as the full channel length, \(L_{\rm h}=L\). Water was used as the working fluid, with \(\rho=1000~\mathrm{kg/m^{3}}\), \(\mu=1.0\times10^{-3}~\mathrm{Pa\,s}\), \(k_{\rm f}=0.6~\mathrm{W/(m\,K)}\), and \(c_{p,\rm f}=4180~\mathrm{J/(kg\,K)}\). A uniform velocity and temperature were prescribed at the inlet, while the outlet pressure was set to zero. No-slip conditions were imposed on all channel walls, which were subjected to a uniform heat flux. The hydraulic and thermal predictions were evaluated using the Darcy friction factor and the mean Nusselt number:
\begin{equation}
f_{\rm D}=\frac{2\Delta pD_{\rm h}}{\rho u_{\rm c}^{\,2}L},
\qquad
\mathrm{Nu}=\frac{\bar{h}D_{\rm h}}{k_{\rm f}},
\label{eq:reference_performance_metrics}
\end{equation}
where \(\bar{h}\) is the mean heat transfer coefficient over the heated walls. For \(\psi\in\{f_{\rm D},\mathrm{Nu}\}\), the relative difference from a reference value was defined as
\begin{equation}
e_{\psi}=\frac{\psi-\psi_{\rm ref}}{\psi_{\rm ref}}\times100\%,
\label{eq:normalized_difference}
\end{equation}
where \(e_{\psi}\) is the percentage relative difference in \(\psi\), and \(\psi_{\rm ref}\) denotes the reference value selected for each comparison.

\subsubsection{Mesh-independence study}
\label{subsubsec:mesh_independence}

Mesh sensitivity was examined at \(\mathrm{Re}=10\) and \(1000\) using structured hexahedral meshes successively refined in the streamwise and cross-sectional directions. These two Reynolds numbers represent different levels of flow and thermal development. The relative difference \(e_{\psi}\) was evaluated using Eq.~\eqref{eq:normalized_difference}, where \(\psi\) denotes the result obtained using the current mesh and \(\psi_{\rm ref}\) denotes that obtained using the preceding coarser mesh.

The mesh-independence results are summarized in Table~\ref{tab:mesh_independence_reference}. The variations in both \(f_{\rm D}\) and \(\mathrm{Nu}\) decreased progressively with increasing mesh resolution. For the \(12\times360\times12\) mesh, the changes relative to the preceding mesh were within \(1.5\%\) at both Reynolds numbers. This mesh was adopted for the subsequent straight-channel calculations.

\subsubsection{Comparison with benchmark solutions}
\label{subsubsec:analytical_comparison}

The accuracy of the mesh-independent model was evaluated by comparing the predicted friction factor and Nusselt number with the developing-flow solutions of Muzychka and Yovanovich~\cite{muzychka2002laminar,muzychka2002laminar2}. Calculations were performed over the range \(10\leq\mathrm{Re}\leq1000\). The analytical values of \(f_{\rm D}\) and \(\mathrm{Nu}\) were obtained from Eqs.~\eqref{eq:sfc_po} and \eqref{eq:nu_developing}, respectively. For this comparison, \(\psi\) and \(\psi_{\rm ref}\) in Eq.~\eqref{eq:normalized_difference} denote the three-dimensional and analytical results, respectively.

Table~\ref{tab:analytical_reference_comparison} compares the numerical and analytical values. Good agreement was obtained throughout the investigated Reynolds-number range, with maximum absolute differences of \(0.82\%\) for \(f_{\rm D}\) and \(0.85\%\) for \(\mathrm{Nu}\). These results confirm that the three-dimensional implementation accurately predicts both the hydraulic resistance and developing convective heat transfer in the straight channel.
\begin{table}[h]
\centering
\caption{Mesh-independence results for the straight-channel benchmark. The signed errors indicate changes relative to the preceding coarser mesh.}
\label{tab:mesh_independence_reference}
\scriptsize
\setlength{\tabcolsep}{3.0pt}
\renewcommand{\arraystretch}{1.5}
\begin{tabular}{ccccccccccc}
\toprule
\(M_x\) & \(M_y\) & \(M_z\) &
\multicolumn{4}{c}{\(\mathrm{Re}=10\)} &
\multicolumn{4}{c}{\(\mathrm{Re}=1000\)} \\
\cmidrule(lr){4-7}
\cmidrule(lr){8-11}
& & &
\(f_{\rm D}\) & \(e_f\) (\%) &
\(\mathrm{Nu}\) & \(e_{\rm Nu}\) (\%) &
\(f_{\rm D}\) & \(e_f\) (\%) &
\(\mathrm{Nu}\) & \(e_{\rm Nu}\) (\%) \\
\midrule
 4 & 120 &  4 & 5.471 & --   & 5.738 & --    & 0.116 & --   & 19.281 & --    \\
 6 & 180 &  6 & 5.638 & 3.04 & 5.584 & -2.69 & 0.119 & 3.14 & 17.576 & -8.84 \\
 8 & 240 &  8 & 5.736 & 1.74 & 5.546 & -0.68 & 0.122 & 2.29 & 16.888 & -3.91 \\
10 & 300 & 10 & 5.801 & 1.15 & 5.533 & -0.24 & 0.124 & 1.64 & 16.519 & -2.18 \\
12 & 360 & 12 & 5.849 & 0.83 & 5.527 & -0.10 & 0.125 & 1.19 & 16.298 & -1.34 \\
14 & 420 & 14 & 5.887 & 0.64 & 5.525 & -0.04 & 0.127 & 0.90 & 16.152 & -0.90 \\
16 & 480 & 16 & 5.917 & 0.51 & 5.524 & -0.02 & 0.127 & 0.71 & 16.050 & -0.63 \\
\bottomrule
\end{tabular}
\end{table}

\begin{table}[h]
\centering
\caption{Comparison of the analytical and three-dimensional results for the straight-channel benchmark.}
\label{tab:analytical_reference_comparison}
\scriptsize
\setlength{\tabcolsep}{5.0pt}
\renewcommand{\arraystretch}{1.5}
\begin{tabular}{ccccccccc}
\toprule
\(\mathrm{Re}\) & \(L^{+}\) & \(x^{*}\) &
\multicolumn{2}{c}{Analytical} &
\multicolumn{2}{c}{3D model} &
\multicolumn{2}{c}{Error (\%)} \\
\cmidrule(lr){4-5}
\cmidrule(lr){6-7}
\cmidrule(lr){8-9}
& & &
\(f_{\rm D}\) & \(\mathrm{Nu}\) &
\(f_{\rm D}\) & \(\mathrm{Nu}\) &
\(e_f\) & \(e_{\rm Nu}\) \\
\midrule
10   & 1.500 & 0.2152 & 5.802 & 4.176  & 5.849 & 4.177  & 0.82  & 0.04 \\
100  & 0.150 & 0.0215 & 0.671 & 6.521  & 0.668 & 6.514  & -0.47 & -0.10 \\
200  & 0.075 & 0.0108 & 0.380 & 8.026  & 0.379 & 8.010  & -0.16 & -0.20 \\
400  & 0.038 & 0.0055 & 0.228 & 10.173 & 0.229 & 10.134 & 0.47  & -0.38 \\
600  & 0.025 & 0.0036 & 0.173 & 11.834 & 0.174 & 11.770 & 0.36  & -0.54 \\
800  & 0.019 & 0.0027 & 0.144 & 13.246 & 0.144 & 13.154 & -0.03 & -0.69 \\
1000 & 0.015 & 0.0022 & 0.126 & 14.499 & 0.125 & 14.376 & -0.41 & -0.85 \\
\bottomrule
\end{tabular}
\end{table}

\subsection{Straight-channel flow and conjugate heat transfer}
\label{subsec:baseline_straight_channel}

\begin{figure}[h]
\centering
\includegraphics[width=0.75\linewidth]{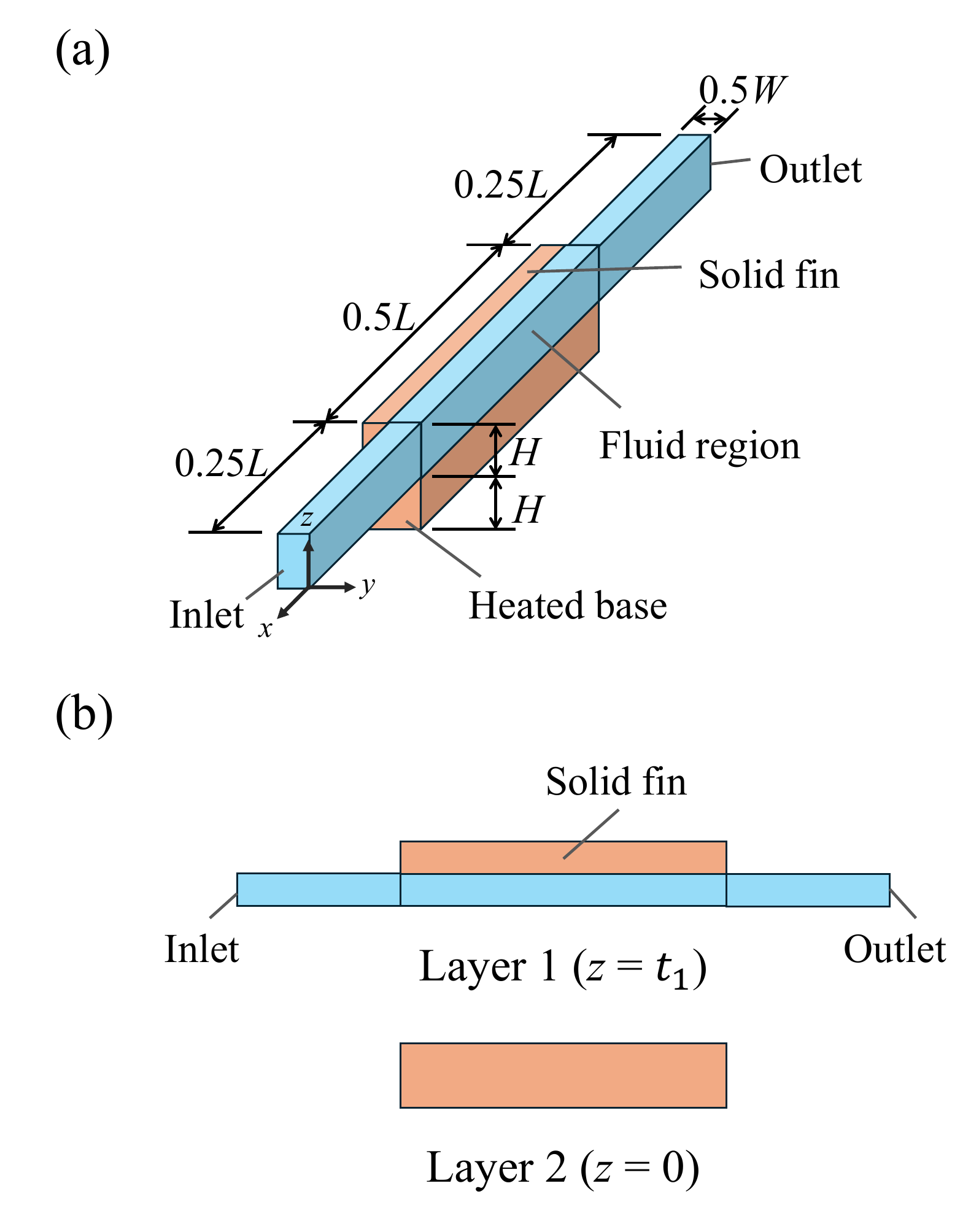}
\caption{Computational domains for the baseline straight-channel conjugate heat-transfer case: (a) symmetry-reduced three-dimensional model and (b) corresponding VTLM representation.}
\label{fig:baseline_straight_channel}
\end{figure}

A straight-channel conjugate heat-transfer case was used as the baseline configuration for calibrating the representative-height closure in Eq.~(\ref{eq:representative_height}). Unlike the fluid-only benchmark in Section~\ref{subsec:reference_model_verification}, this configuration includes heat conduction through a solid fin and a heated base. Fig.~\ref{fig:baseline_straight_channel}(a) shows the three-dimensional domain, while Fig.~\ref{fig:baseline_straight_channel}(b) shows its two-layer representation. The channel consists of inlet and outlet sections of length \(0.25L\) and a central section of length \(0.5L\) containing the fin and heated base. The wall-friction closure was evaluated separately for these streamwise sections, with \(L_{\rm h}=0.25L\) for the inlet and outlet sections and \(L_{\rm h}=0.5L\) for the central section. Unless otherwise stated, the \(L^{+}\) values reported in the subsequent conjugate heat-transfer assessments refer to the central section and are evaluated using \(L_{\rm h}=0.5L\). In the symmetry-reduced domain, both the flow passage and the fin have a transverse width of \(0.5W\).

The channel length and width were \(L=45.0~\mathrm{mm}\) and \(W=3.0~\mathrm{mm}\), respectively. The substrate thickness was fixed at \(t_2=3.0~\mathrm{mm}\) for all conjugate heat-transfer cases considered in this study. The water properties were identical to those used in Section~\ref{subsec:reference_model_verification}, and copper with \(k_{\rm s}=386~\mathrm{W/(m\,K)}\) was assigned to the fin and heated base. The inlet temperature was fixed at \(T_{\rm in}=293.15~\mathrm{K}\), and a uniform heat flux of \(q''_{\rm bt}=50~\mathrm{kW/m^2}\) was applied at the bottom surface. The Reynolds number was maintained at \(\mathrm{Re}=400\). The geometric aspect ratio was defined as \(\mathrm{AR}=H/W\) and varied from \(0.2\) to \(5.0\) by changing \(H\) while keeping \(W\) fixed.

The hydraulic comparison was performed using the Darcy friction factor defined in Eq.~\eqref{eq:reference_performance_metrics}, with the relative difference \(e_f\) evaluated using Eq.~\eqref{eq:normalized_difference}. For the thermal comparison, the area-averaged bottom-surface temperature \(\overline{T}_{\rm bt}\) was adopted as the primary thermal metric, consistent with previous two-layer models for microchannel heat sinks~\cite{yan2019topology,zeng2019topology}. To avoid a temperature-relative error that depends on the chosen temperature reference, the thermal discrepancy was defined as the signed difference
\[
\Delta T_{\rm bt}
=
\overline{T}_{\rm bt}^{\rm VTLM}
-
\overline{T}_{\rm bt}^{\rm 3D}.
\]

The calibration results are summarized in Table~\ref{tab:baseline_straight_channel}. Across the baseline dataset, the hydraulic discrepancy remains within \(4.62\%\), while the signed bottom-temperature difference ranges from \(-1.46\) to \(1.32~^\circ\mathrm{C}\). At low aspect ratios, the VTLM slightly overpredicts \(f_{\rm D}\) and underpredicts \(\overline{T}_{\rm bt}\). The bottom-temperature difference changes sign as \(\mathrm{AR}\) increases, whereas the hydraulic prediction remains slightly higher than the three-dimensional value in all cases. The parameters in Eq.~\eqref{eq:representative_height} were determined from this baseline dataset and retained without readjustment in the remaining cases. Accordingly, these baseline cases are treated as calibration cases and are not included in the subsequent assessment of the VTLM prediction accuracy.

\begin{table}[h]
\centering
\caption{Comparison of the VTLM and three-dimensional reference results for the baseline straight-channel conjugate heat-transfer cases used for calibration.}
\label{tab:baseline_straight_channel}
\scriptsize
\setlength{\tabcolsep}{5pt}
\renewcommand{\arraystretch}{1.5}
\begin{tabular}{ccccccc}
\toprule
\(\mathrm{AR}\) &
\(f_{\rm D}^{\rm 3D}\) &
\(\overline{T}_{\rm bt}^{\rm 3D}\) (\(^\circ\mathrm{C}\)) &
\(f_{\rm D}^{\rm VTLM}\) &
\(\overline{T}_{\rm bt}^{\rm VTLM}\) (\(^\circ\mathrm{C}\)) &
\(e_f\) (\%) &
\(\Delta T_{\rm bt}\) (\(^\circ\mathrm{C}\)) \\
\midrule
0.2 & 0.206 & 36.49 & 0.213 & 35.03 & 3.44 & -1.46 \\
0.3 & 0.202 & 38.38 & 0.207 & 37.30 & 2.30 & -1.08 \\
0.4 & 0.202 & 39.08 & 0.204 & 38.46 & 1.03 & -0.62 \\
0.5 & 0.205 & 39.10 & 0.206 & 38.77 & 0.46 & -0.33 \\
1.0 & 0.225 & 36.34 & 0.230 & 36.85 & 2.10 &  0.51 \\
2.0 & 0.259 & 31.55 & 0.268 & 32.67 & 3.50 &  1.12 \\
3.0 & 0.276 & 29.04 & 0.288 & 30.34 & 4.28 &  1.30 \\
4.0 & 0.288 & 27.63 & 0.301 & 28.95 & 4.55 &  1.32 \\
5.0 & 0.295 & 26.77 & 0.309 & 28.04 & 4.62 &  1.27 \\
\bottomrule
\end{tabular}
\end{table}

\begin{table}[!tb]
\centering
\caption{Comparison of the VTLM and existing two-layer models with the three-dimensional reference results at representative channel aspect ratios. The bottom-surface temperature \(T_{\rm bt}\) is given in \(^\circ\mathrm{C}\).}
\label{tab:model_comparison}
\scriptsize
\renewcommand{\arraystretch}{1.5}
\resizebox{\columnwidth}{!}{%
\begin{tabular}{ccccccccc}
\toprule
& \multicolumn{2}{c}{3D reference}
& \multicolumn{2}{c}{VTLM}
& \multicolumn{2}{c}{Zeng and Lee}
& \multicolumn{2}{c}{Yan et al.} \\
\cmidrule(lr){2-3}
\cmidrule(lr){4-5}
\cmidrule(lr){6-7}
\cmidrule(lr){8-9}
\(\mathrm{AR}\)
& \(f_{\rm D}\)
& \(\overline{T}_{\rm bt}\)
& \(f_{\rm D}\)
& \(\overline{T}_{\rm bt}\)
& \(f_{\rm D}\)
& \(\overline{T}_{\rm bt}\)
& \(f_{\rm D}\)
& \(\overline{T}_{\rm bt}\) \\
\midrule
0.2
& 0.206 & 36.49
& 0.213 & 35.03
& 0.019 & 46.70
& 0.163 & 46.89 \\
1.0
& 0.225 & 36.34
& 0.230 & 36.85
& 0.108 & 40.60
& 0.157 & 43.70 \\
5.0
& 0.295 & 26.77
& 0.309 & 28.04
& 0.254 & 27.90
& 0.250 & 28.65 \\
\bottomrule
\end{tabular}%
}
\end{table}

\begin{figure}[h]
\centering
\includegraphics[width=0.82\linewidth]{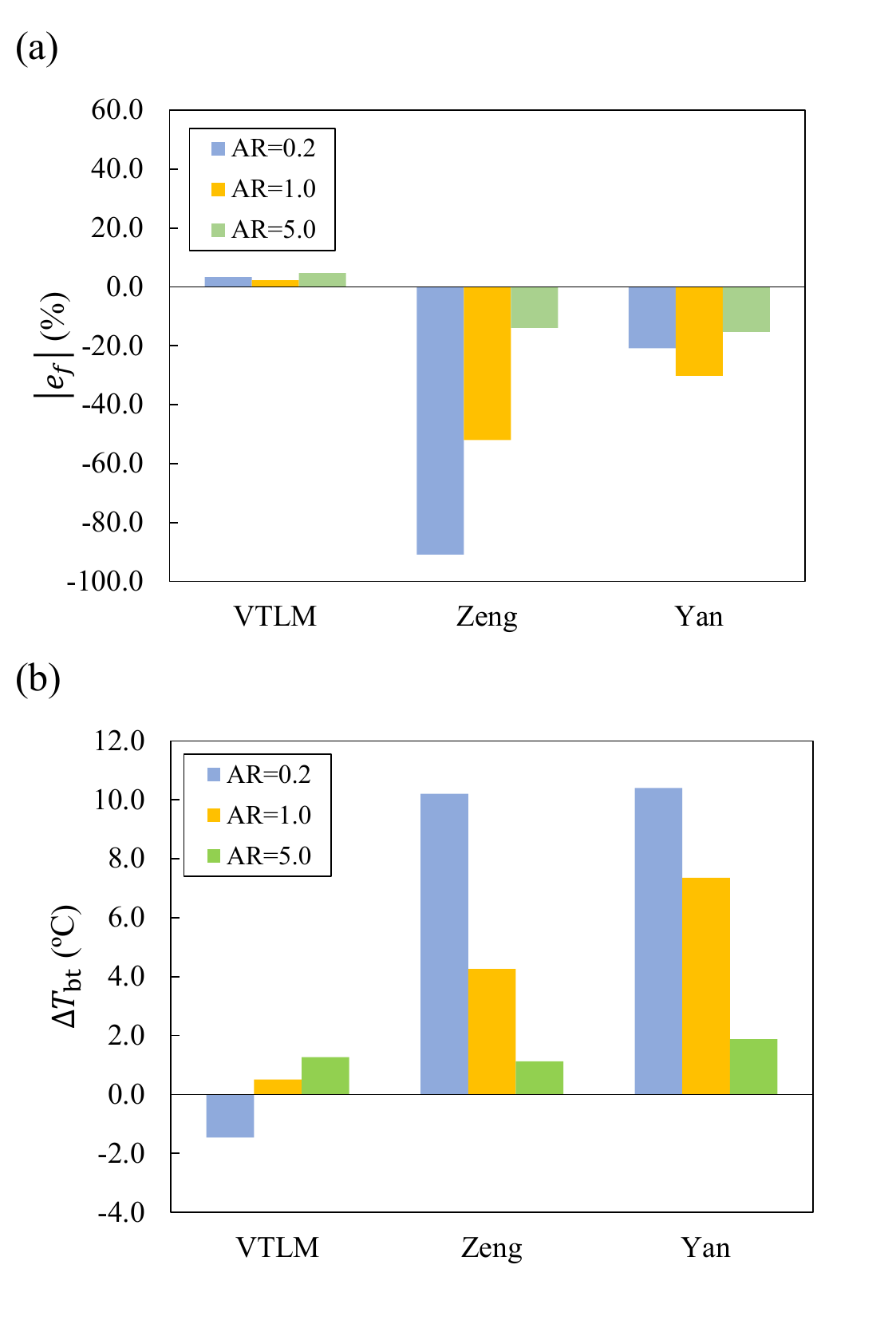}
\caption{Prediction differences from the three-dimensional reference results for the VTLM and existing two-layer models at \(\mathrm{AR}=0.2\), 1.0, and 5.0: (a) relative difference in the Darcy friction factor and (b) area-averaged bottom-temperature difference.}
\label{fig:model_comparison_errors}
\end{figure}

To further compare the behavior of the VTLM with existing reduced-dimensional approaches, the two-layer models of Zeng and Lee~\cite{zeng2019topology} and Yan et al.~\cite{yan2019topology} were evaluated at representative channel aspect ratios. The predicted values are summarized in Table~\ref{tab:model_comparison}, and the corresponding hydraulic and thermal prediction differences from the three-dimensional reference results are shown in Fig.~\ref{fig:model_comparison_errors}. Because these baseline configurations were used only to calibrate the representative-height thermal closure, the hydraulic comparison remains independent of that calibration. Therefore, the thermal comparison is included to illustrate the behavior of the different reduced-dimensional formulations under the same baseline conditions rather than to provide an independent assessment of the VTLM.

The hydraulic comparison in Fig.~\ref{fig:model_comparison_errors}(a) shows the substantial differences among the models. At \(\mathrm{AR}=0.2\), the VTLM predicts the Darcy friction factor within \(3.40\%\) of the three-dimensional result, whereas the models of Zeng and Lee and Yan et al. underpredict it by \(90.78\%\) and \(20.87\%\), respectively. The large discrepancy of the Zeng and Lee model is consistent with its omission of the friction associated with the unresolved top and bottom walls, whose contribution becomes particularly important at low aspect ratios. The Yan et al. model also shows appreciable underprediction because its plane-flow formulation does not account for hydrodynamic development in the height direction. By contrast, the VTLM incorporates both effects through its hydraulic closure. As \(\mathrm{AR}\) increases, the errors of the existing models decrease substantially because the relative contribution of the top and bottom wall friction becomes smaller.

\begin{table}[h]
\centering
\caption{Computational cost of the VTLM and three-dimensional reference model for representative straight-channel conjugate heat-transfer cases.}
\label{tab:computational_cost}
\scriptsize
\setlength{\tabcolsep}{8pt}
\renewcommand{\arraystretch}{1.4}
\begin{tabular}{cccccc}
\toprule
\(\mathrm{AR}\) &
3D DOFs &
VTLM DOFs &
\(t_{\rm 3D}\) (s) &
\(t_{\rm VTLM}\) (s) &
Speedup \\
\midrule
0.2 & 749525  & 25469 & 137 & 7 & 19.6 \\
1.0 & 749525  & 25469 & 138 & 4 & 34.5 \\
5.0 & 3373577 & 25469 & 682 & 4 & 170.5 \\
\bottomrule
\end{tabular}
\end{table}

\begin{table*}[t]
\centering
\caption{Comparison of the VTLM and three-dimensional reference results for uniform transverse fin arrays at \(\mathrm{AR}=1\).}
\label{tab:uniform_transverse_ar1}
\scriptsize
\setlength{\tabcolsep}{3.5pt}
\renewcommand{\arraystretch}{1.5}
\begin{tabular*}{0.8\textwidth}{@{\extracolsep{\fill}}cccccccccccc@{}}
\toprule
Case & \(\mathrm{Re}\) & \(\mathrm{AR}\) & \(L^{+}\) & \(N_y\) & \(\lambda_y\) &
\(f_{\rm D}^{\rm 3D}\) & \(f_{\rm D}^{\rm VTLM}\) & \(e_f\) (\%) &
\(\overline{T}_{\rm bt}^{\rm 3D}\) (\(^\circ\mathrm{C}\)) &
\(\overline{T}_{\rm bt}^{\rm VTLM}\) (\(^\circ\mathrm{C}\)) &
\(\Delta T_{\rm bt}\) (\(^\circ\mathrm{C}\)) \\
\midrule
UT1  & 100 & 1.0 & 0.0750 & 1 & 0.10 & 1.338 & 1.374 &  2.72 & 32.43 & 32.72 &  0.29 \\
UT2  & 100 & 1.0 & 0.0750 & 1 & 0.60 & 9.517 & 9.223 & -3.09 & 28.05 & 28.63 &  0.58 \\
UT3  & 100 & 1.0 & 0.0750 & 3 & 0.05 & 3.994 & 3.856 & -3.47 & 25.05 & 25.63 &  0.58 \\
UT4  & 100 & 1.0 & 0.0750 & 3 & 0.10 & 6.767 & 6.502 & -3.93 & 24.40 & 24.99 &  0.59 \\
UT5  & 100 & 1.0 & 0.0750 & 5 & 0.05 & 11.692 & 11.105 & -5.02 & 23.41 & 23.89 &  0.48 \\
UT6  & 100 & 1.0 & 0.0750 & 5 & 0.10 & 37.903 & 36.045 & -4.90 & 22.89 & 23.32 &  0.43 \\
\midrule
UT7  & 500 & 1.0 & 0.0150 & 1 & 0.10 & 0.342 & 0.381 & 11.49 & 26.67 & 27.35 &  0.68 \\
UT8  & 500 & 1.0 & 0.0150 & 1 & 0.40 & 0.861 & 0.857 & -0.47 & 24.94 & 25.84 &  0.90 \\
UT9  & 500 & 1.0 & 0.0150 & 3 & 0.05 & 0.890 & 0.898 &  0.85 & 23.00 & 23.50 &  0.50 \\
UT10 & 500 & 1.0 & 0.0150 & 5 & 0.10 & 8.221 & 7.668 & -6.72 & 21.37 & 21.75 &  0.38 \\
\midrule
UT11 & 500 & 1.0 & 0.0225 & 1 & 0.10 & 0.310 & 0.341 &  9.84 & 25.26 & 25.74 &  0.48 \\
UT12 & 500 & 1.0 & 0.0450 & 1 & 0.10 & 0.276 & 0.293 &  6.20 & 23.59 & 23.65 &  0.06 \\
UT13 & 100 & 1.0 & 0.2250 & 1 & 0.10 & 1.216 & 1.213 & -0.29 & 26.56 & 26.49 & -0.07 \\
UT14 & 100 & 1.0 & 0.0150 & 1 & 0.10 & 1.911 & 2.041 &  6.82 & 53.69 & 56.21 &  2.52 \\
\bottomrule
\end{tabular*}
\end{table*}

The thermal comparison in Fig.~\ref{fig:model_comparison_errors}(b) shows a similar reduction in model differences with increasing aspect ratio. At \(\mathrm{AR}=0.2\), the VTLM gives a bottom-temperature difference of \(-1.46~^\circ\mathrm{C}\), whereas the models of Zeng and Lee and Yan et al. overpredict the bottom temperature by \(10.21~^\circ\mathrm{C}\) and \(10.40~^\circ\mathrm{C}\), respectively. The larger discrepancies of the latter are associated with their simplified treatment of the height-direction thermal behavior, using either an effective thermal-resistance representation or a plane-flow-based formulation. In contrast, the VTLM evaluates its thermal closures from the representative geometry and development condition. At \(\mathrm{AR}=1.0\), the VTLM gives a bottom-temperature difference of only \(0.51~^\circ\mathrm{C}\), compared with \(4.26~^\circ\mathrm{C}\) and \(7.36~^\circ\mathrm{C}\) for the Zeng and Lee and Yan et al. models, respectively. At \(\mathrm{AR}=5.0\), all three models give differences below \(2~^\circ\mathrm{C}\), indicating that the discrepancies among the reduced-dimensional formulations become less pronounced as the aspect ratio increases.

The computational cost of the VTLM was also compared with that of the three-dimensional reference model using the representative straight-channel cases. The solution times reported in Table~\ref{tab:computational_cost} correspond to the stationary solution process. The VTLM substantially reduces the computational cost relative to the three-dimensional model, with solution-time speedups ranging from approximately 20 to 171 over the examined aspect ratios. The computational advantage becomes more pronounced at larger aspect ratios, as the VTLM avoids explicit discretization in the channel-height direction.

\subsection{Fin arrays in a straight channel}
\label{Fin arrays in a straight channel}

\begin{table*}[t]
\centering
\caption{Comparison of the VTLM and three-dimensional reference results for uniform transverse fin arrays at \(\mathrm{AR}=0.2\) and \(5.0\).}
\label{tab:uniform_transverse_ar_effects}
\scriptsize
\setlength{\tabcolsep}{3.5pt}
\renewcommand{\arraystretch}{1.5}
\begin{tabular*}{0.8\textwidth}{@{\extracolsep{\fill}}cccccccccccc@{}}
\toprule
Case & \(\mathrm{Re}\) & \(\mathrm{AR}\) & \(L^{+}\) & \(N_y\) & \(\lambda_y\) &
\(f_{\rm D}^{\rm 3D}\) & \(f_{\rm D}^{\rm VTLM}\) & \(e_f\) (\%) &
\(\overline{T}_{\rm bt}^{\rm 3D}\) (\(^\circ\mathrm{C}\)) &
\(\overline{T}_{\rm bt}^{\rm VTLM}\) (\(^\circ\mathrm{C}\)) &
\(\Delta T_{\rm bt}\) (\(^\circ\mathrm{C}\)) \\
\midrule
UT15 & 100 & 0.2 & 0.225 & 1 & 0.20 & 1.016 & 1.051 &  3.44 & 34.65 & 33.21 & -1.44 \\
UT16 & 100 & 0.2 & 0.225 & 1 & 0.60 & 2.586 & 2.509 & -2.98 & 35.41 & 34.73 & -0.68 \\
UT17 & 500 & 0.2 & 0.045 & 1 & 0.20 & 0.291 & 0.260 & -10.65 & 27.24 & 26.95 & -0.29 \\
UT18 & 500 & 0.2 & 0.045 & 1 & 0.60 & 0.794 & 0.700 & -11.84 & 26.20 & 30.41 &  4.21 \\
UT19 & 100 & 0.2 & 0.225 & 3 & 0.05 & 1.263 & 1.273 &  0.79 & 29.23 & 27.96 & -1.27 \\
UT20 & 500 & 0.2 & 0.045 & 3 & 0.05 & 0.284 & 0.306 &  7.75 & 24.58 & 24.23 & -0.35 \\
UT21 & 500 & 0.2 & 0.045 & 3 & 0.15 & 0.631 & 0.618 & -2.06 & 23.97 & 23.84 & -0.13 \\
UT22 & 500 & 0.2 & 0.045 & 5 & 0.075 & 0.766 & 0.725 & -5.35 & 22.83 & 22.60 & -0.23 \\
\midrule
UT23 & 100 & 5.0 & 0.045 & 1 & 0.20 & 3.360 & 3.422 &  1.85 & 26.03 & 26.82 &  0.79 \\
UT24 & 100 & 5.0 & 0.045 & 1 & 0.40 & 7.248 & 7.260 &  0.17 & 24.33 & 25.24 &  0.91 \\
UT25 & 500 & 5.0 & 0.009 & 1 & 0.20 & 0.784 & 0.806 &  2.81 & 24.08 & 24.71 &  0.63 \\
UT26 & 500 & 5.0 & 0.009 & 1 & 0.40 & 1.708 & 1.693 & -0.88 & 22.85 & 23.60 &  0.75 \\
UT27 & 500 & 5.0 & 0.009 & 1 & 0.60 & 5.483 & 5.320 & -2.97 & 22.15 & 22.85 &  0.70 \\
UT28 & 100 & 5.0 & 0.045 & 3 & 0.05 & 9.405 & 9.397 & -0.09 & 24.12 & 24.22 &  0.10 \\
UT29 & 500 & 5.0 & 0.009 & 3 & 0.05 & 2.019 & 2.010 & -0.45 & 22.80 & 23.11 &  0.31 \\
UT30 & 500 & 5.0 & 0.009 & 3 & 0.15 & 7.090 & 7.038 & -0.73 & 21.67 & 21.98 &  0.31 \\
UT31 & 500 & 5.0 & 0.009 & 5 & 0.075 & 10.478 & 10.601 &  1.17 & 21.51 & 21.74 &  0.23 \\
\bottomrule
\end{tabular*}
\end{table*}

\begin{table*}[!tb]
\centering
\caption{Comparison of the VTLM and three-dimensional reference results for uniform streamwise segmented fin arrays.}
\label{tab:uniform_streamwise_segmented_fin_arrays}
\scriptsize
\setlength{\tabcolsep}{3.5pt}
\renewcommand{\arraystretch}{1.6}
\begin{tabular*}{0.8\textwidth}{@{\extracolsep{\fill}}cccccccccccc@{}}
\toprule
Case & \(\mathrm{Re}\) & \(\mathrm{AR}\) & \(L^{+}\) & \(N_x\) & \(\lambda_x\) &
\(f_{\rm D}^{\rm 3D}\) & \(f_{\rm D}^{\rm VTLM}\) & \(e_f\) (\%) &
\(\overline{T}_{\rm bt}^{\rm 3D}\) (\(^\circ\mathrm{C}\)) &
\(\overline{T}_{\rm bt}^{\rm VTLM}\) (\(^\circ\mathrm{C}\)) &
\(\Delta T_{\rm bt}\) (\(^\circ\mathrm{C}\)) \\
\midrule
S1  & 100 & 1.0 & 0.075 & 2 & 0.225 & 1.669 & 1.687 & 1.06 & 31.68 & 32.06 &  0.38 \\
S2  & 100 & 1.0 & 0.075 & 2 & 0.125 & 1.325 & 1.352 & 2.04 & 33.60 & 34.15 &  0.55 \\
S3  & 500 & 1.0 & 0.015 & 2 & 0.225 & 0.432 & 0.463 & 7.17 & 25.88 & 26.77 &  0.89 \\
S4  & 500 & 1.0 & 0.015 & 2 & 0.125 & 0.379 & 0.407 & 7.38 & 26.32 & 27.75 &  1.43 \\
\midrule
S5  & 100 & 1.0 & 0.075 & 3 & 0.133 & 1.638 & 1.653 & 0.93 & 31.83 & 32.17 &  0.34 \\
S6  & 100 & 1.0 & 0.075 & 3 & 0.067 & 1.312 & 1.332 & 1.53 & 33.63 & 34.31 &  0.68 \\
S7  & 500 & 1.0 & 0.015 & 3 & 0.133 & 0.429 & 0.459 & 7.06 & 25.84 & 26.78 &  0.94 \\
S8  & 500 & 1.0 & 0.015 & 3 & 0.067 & 0.383 & 0.409 & 6.71 & 26.02 & 27.72 &  1.70 \\
\midrule
S9  & 100 & 0.2 & 0.225 & 2 & 0.175 & 0.970 & 1.000 & 3.14 & 34.41 & 32.87 & -1.54 \\
S10 & 100 & 0.2 & 0.225 & 3 & 0.10 & 0.965 & 0.992 & 2.80 & 34.25 & 32.61 & -1.64 \\
S11 & 500 & 0.2 & 0.045 & 2 & 0.175 & 0.231 & 0.253 & 9.62 & 26.67 & 26.58 & -0.09 \\
S12 & 500 & 0.2 & 0.045 & 3 & 0.10 & 0.232 & 0.254 & 9.33 & 26.31 & 26.32 &  0.01 \\
\midrule
S13 & 100 & 5.0 & 0.045 & 2 & 0.175 & 2.844 & 2.904 & 2.10 & 27.35 & 27.87 &  0.52 \\
S14 & 100 & 5.0 & 0.045 & 3 & 0.10 & 2.775 & 2.832 & 2.05 & 27.77 & 28.13 &  0.36 \\
S15 & 500 & 5.0 & 0.009 & 2 & 0.175 & 0.717 & 0.732 & 2.04 & 24.71 & 25.35 &  0.64 \\
S16 & 500 & 5.0 & 0.009 & 3 & 0.10 & 0.718 & 0.729 & 1.51 & 24.84 & 25.51 &  0.67 \\
\bottomrule
\end{tabular*}
\end{table*}

The straight-channel assessment was extended from the baseline configuration shown in Fig.~\ref{fig:baseline_straight_channel} to channels containing uniformly arranged fin arrays. The fins were placed within the central heated section of length \(L_{\rm fp}=0.5L\) and were assigned identical rectangular footprints. The streamwise fin length and transverse fin width are denoted by \(L_{\rm fin}\) and \(W_{\rm fin}\), respectively, and their dimensionless values are defined as

\begin{equation}
\lambda_x=\frac{L_{\rm fin}}{L},
\qquad
\lambda_y=\frac{W_{\rm fin}}{W}.
\label{eq:dimensionless_fin_dimensions}
\end{equation}

The numbers of fins in the streamwise and transverse directions are denoted by \(N_x\) and \(N_y\), respectively. In each direction containing multiple fins, the fins were uniformly spaced. The channel dimensions, material properties, and boundary conditions were the same as those used for the straight-channel conjugate heat-transfer case. The VTLM predictions were compared with the corresponding three-dimensional results using \(f_{\rm D}\) and \(\overline{T}_{\rm bt}\). The hydraulic discrepancy was evaluated using the relative difference \(e_f\), while the thermal discrepancy was quantified by the signed bottom-temperature difference.

\begin{figure}[h]
\centering
\includegraphics[width=0.6\linewidth]{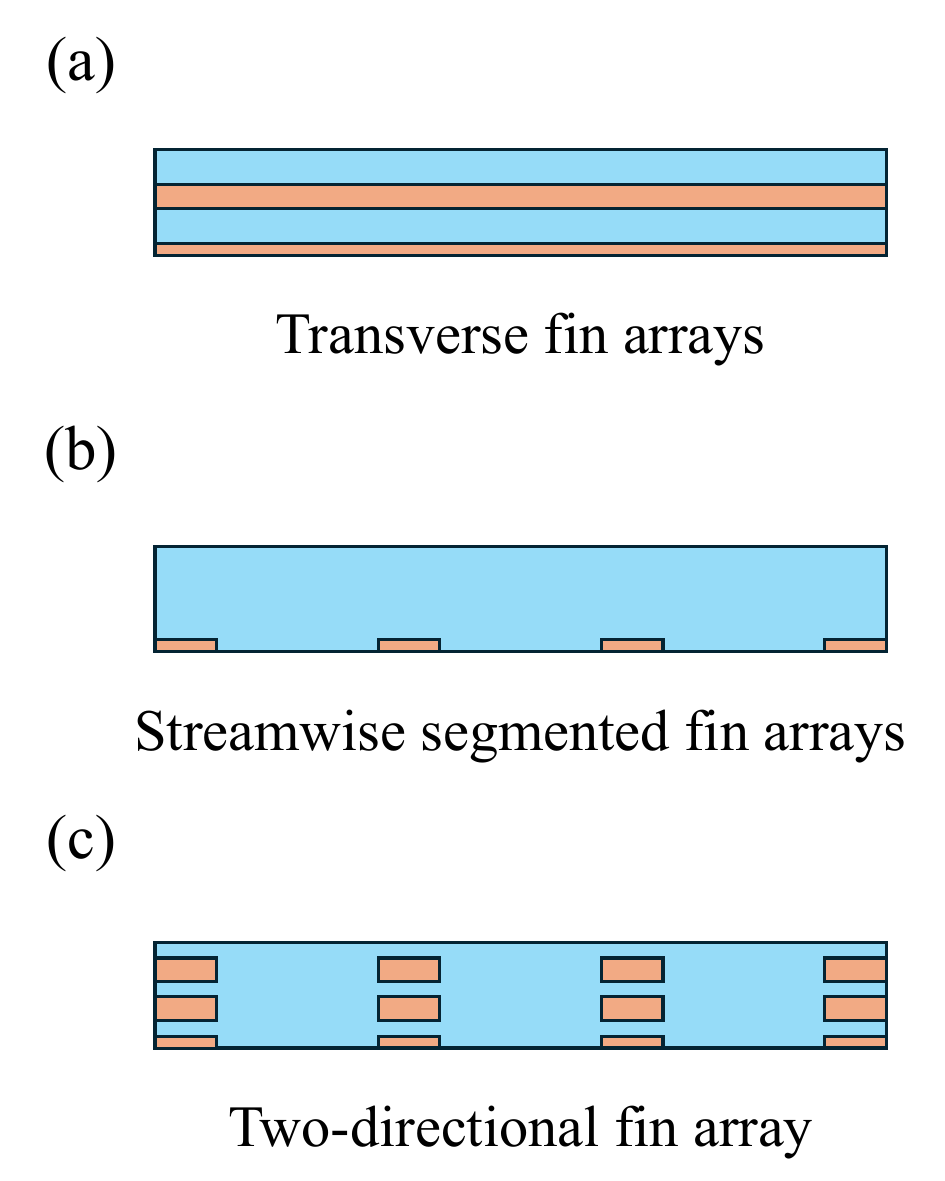}
\caption{Uniform fin-array groups considered in the straight-channel assessment. Only the central heated section is shown: (a) transverse fin arrays with \(N_x=1\) and \(N_y>1\), (b) streamwise segmented fin arrays with \(N_x>1\) and \(N_y=1\), and (c) two-directional fin arrays with \(N_x>1\) and \(N_y>1\).}
\label{fig:uniform_fin_array_groups}
\end{figure}

The uniform fin-array cases were divided into the three groups illustrated in Fig.~\ref{fig:uniform_fin_array_groups}. Transverse fin arrays have a single streamwise fin section, \(N_x=1\) and \(N_y>1\), while the transverse fin number and width are varied. Streamwise segmented fin arrays have \(N_x>1\) and \(N_y=1\), allowing the effects of segment number and streamwise fin length to be examined. Two-directional fin arrays have \(N_x>1\) and \(N_y>1\), providing configurations in which the flow passage is subdivided in both in-plane directions.

\subsubsection{Uniform transverse fin arrays}
\label{subsec:uniform_transverse_fin_arrays}

Uniform transverse fin arrays were examined using the configuration shown in Fig.~\ref{fig:uniform_fin_array_groups}(a). For these cases, \(N_x=1\) and \(\lambda_x=0.5\), so that each fin extends over the full heated-section length. The transverse arrangement was varied through the number of fins \(N_y\) and the normalized fin width \(\lambda_y\). Cases UT11--UT14 vary the hydraulic diameter at fixed \(\mathrm{AR}=1\) and with the same normalized single-fin configuration to examine the influence of \(L^+\) separately from that of the Reynolds number. The uniform transverse fin-array cases and the corresponding results are summarized in Table~\ref{tab:uniform_transverse_ar1} for \(\mathrm{AR}=1\) and in Table~\ref{tab:uniform_transverse_ar_effects} for \(\mathrm{AR}=0.2\) and \(5.0\).

For \(\mathrm{AR}=1\), the dominant change in prediction accuracy is associated with the flow-development condition. At \(L^{+}=0.075\), the maximum absolute hydraulic error and bottom-temperature difference are \(5.02\%\) and \(0.59~^\circ\mathrm{C}\), respectively. Reducing \(L^{+}\) to \(0.015\) increases the hydraulic sensitivity, particularly for the weakly obstructed single-fin configuration UT7, whereas increasing the fin width or subdividing the solid region among multiple fins generally reduces this bias. The bottom-temperature differences remain within \(0.90~^\circ\mathrm{C}\) for these \(\mathrm{AR}=1\) configurations.

The aspect ratio and blockage distribution modify this dependence on \(L^{+}\). At \(\mathrm{AR}=0.2\), combining a low flow-development length with a large localized blockage produces the largest discrepancy, as shown by Case UT18, with a hydraulic error of \(-11.84\%\) and a bottom-temperature difference of \(4.21~^\circ\mathrm{C}\). Configurations with multiple narrower fins exhibit substantially smaller temperature differences than the single wide-fin case. By contrast, the \(\mathrm{AR}=5.0\) cases remain thermally robust, with absolute bottom-temperature differences within \(0.91~^\circ\mathrm{C}\), even at \(L^{+}=0.009\). These comparisons show that \(L^{+}\) governs the overall hydraulic accuracy, while the aspect ratio and localization of the transverse blockage influence the remaining temperature discrepancy.

\subsubsection{Uniform streamwise segmented fin arrays}
\label{subsec:uniform_streamwise_segmented_fin_arrays}

Uniform streamwise segmented fin arrays were examined using the configuration shown in Fig.~\ref{fig:uniform_fin_array_groups}(b). The transverse fin number was fixed at \(N_y=1\) and \(\lambda_y=0.2\), while the full-length fin was divided into \(N_x=2\) or \(3\) equal segments distributed uniformly within the heated section. The fin-array cases and corresponding results are summarized in Table~\ref{tab:uniform_streamwise_segmented_fin_arrays}.

Across the streamwise segmented arrays, reducing \(L^{+}\) has a greater influence on the VTLM accuracy than varying the segment number or length. At \(\mathrm{AR}=1\), the maximum hydraulic error increases from \(2.04\%\) at \(L^{+}=0.075\) to \(7.38\%\) at \(L^{+}=0.015\), while the absolute bottom-temperature differences remain within \(1.70~^\circ\mathrm{C}\). Paired cases with \(N_x=2\) and \(3\) give similar results when their segment dimensions are comparable, indicating that additional streamwise segmentation does not cause a systematic deterioration.

The aspect-ratio comparisons support the same interpretation. At \(\mathrm{AR}=0.2\), reducing \(L^{+}\) from \(0.225\) to \(0.045\) increases the hydraulic error to approximately \(9.5\%\), while the absolute bottom-temperature differences remain within \(1.64~^\circ\mathrm{C}\). At \(\mathrm{AR}=5.0\), the hydraulic errors remain within \(2.10\%\), and the absolute bottom-temperature differences remain within \(0.67~^\circ\mathrm{C}\), even at \(L^{+}=0.009\). Thus, the prediction differences are governed primarily by the flow-development condition and channel aspect ratio, with no distinct limitation attributable to streamwise segmentation.

\begin{table*}[!tb]
\centering
\caption{Comparison of the VTLM and three-dimensional reference results for uniform two-directional fin arrays at \(\mathrm{AR}=1\).}
\label{tab:uniform_two_directional_ar1}
\scriptsize
\setlength{\tabcolsep}{2.2pt}
\renewcommand{\arraystretch}{1.6}
\begin{tabular*}{0.8\textwidth}{@{\extracolsep{\fill}}cccccccccccccc@{}}
\toprule
Case & \(\mathrm{Re}\) & \(\mathrm{AR}\) & \(L^{+}\) & \(N_x\) & \(N_y\) & \(\lambda_x\) & \(\lambda_y\) & \(f_{\rm D}^{\rm 3D}\) & \(f_{\rm D}^{\rm VTLM}\) & \(e_f\) (\%) & \(\overline{T}_{\rm bt}^{\rm 3D}\) (\(^\circ\mathrm{C}\)) & \(\overline{T}_{\rm bt}^{\rm VTLM}\) (\(^\circ\mathrm{C}\)) & \(\Delta T_{\rm bt}\) (\(^\circ\mathrm{C}\)) \\
\midrule
D1  & 100 & 1.0 & 0.075 & 2 & 3 & 0.225 & 0.100 & 6.361 & 6.102 & -4.09 & 24.50 & 25.05 & 0.55 \\
D2  & 100 & 1.0 & 0.075 & 2 & 3 & 0.075 & 0.050 & 1.831 & 1.831 &  0.02 & 29.42 & 30.08 & 0.66 \\
D3  & 100 & 1.0 & 0.075 & 5 & 3 & 0.060 & 0.100 & 5.137 & 4.897 & -4.68 & 24.91 & 25.56 & 0.65 \\
D4  & 100 & 1.0 & 0.075 & 5 & 3 & 0.020 & 0.100 & 2.785 & 2.672 & -4.06 & 27.13 & 28.40 & 1.27 \\
D5  & 100 & 1.0 & 0.075 & 5 & 5 & 0.060 & 0.050 & 8.116 & 7.704 & -5.08 & 23.92 & 24.36 & 0.44 \\
D6  & 100 & 1.0 & 0.075 & 9 & 3 & 0.033 & 0.100 & 5.614 & 5.383 & -4.12 & 24.91 & 25.29 & 0.38 \\
\midrule
D7  & 500 & 1.0 & 0.015 & 2 & 3 & 0.225 & 0.100 & 1.450 & 1.420 & -2.04 & 22.40 & 22.95 & 0.55 \\
D8  & 500 & 1.0 & 0.015 & 2 & 3 & 0.075 & 0.050 & 0.496 & 0.516 &  4.10 & 25.41 & 26.16 & 0.75 \\
D9  & 500 & 1.0 & 0.015 & 5 & 3 & 0.060 & 0.100 & 1.322 & 1.273 & -3.69 & 22.44 & 23.13 & 0.69 \\
D10 & 500 & 1.0 & 0.015 & 5 & 3 & 0.020 & 0.100 & 0.944 & 0.886 & -6.12 & 23.40 & 24.68 & 1.28 \\
D11 & 500 & 1.0 & 0.015 & 5 & 5 & 0.060 & 0.050 & 2.013 & 1.875 & -6.86 & 22.02 & 22.48 & 0.46 \\
D12 & 500 & 1.0 & 0.015 & 9 & 3 & 0.033 & 0.100 & 1.351 & 1.347 & -0.34 & 22.60 & 22.95 & 0.35 \\
\bottomrule
\end{tabular*}
\end{table*}

\begin{table*}[!tb]
\centering
\caption{Comparison of the VTLM and three-dimensional reference results for uniform two-directional fin arrays at \(\mathrm{AR}=0.2\) and \(5.0\).}
\label{tab:uniform_two_directional_ar_effects}
\scriptsize
\setlength{\tabcolsep}{2.2pt}
\renewcommand{\arraystretch}{1.6}
\begin{tabular*}{0.8\textwidth}{@{\extracolsep{\fill}}cccccccccccccc@{}}
\toprule
Case & \(\mathrm{Re}\) & \(\mathrm{AR}\) & \(L^{+}\) & \(N_x\) & \(N_y\) & \(\lambda_x\) & \(\lambda_y\) & \(f_{\rm D}^{\rm 3D}\) & \(f_{\rm D}^{\rm VTLM}\) & \(e_f\) (\%) & \(\overline{T}_{\rm bt}^{\rm 3D}\) (\(^\circ\mathrm{C}\)) & \(\overline{T}_{\rm bt}^{\rm VTLM}\) (\(^\circ\mathrm{C}\)) & \(\Delta T_{\rm bt}\) (\(^\circ\mathrm{C}\)) \\
\midrule
D13 & 500 & 0.2 & 0.045 & 2 & 3 & 0.225 & 0.100 & 0.392 & 0.401 &  2.23 & 24.06 & 23.69 & -0.37 \\
D14 & 500 & 0.2 & 0.045 & 2 & 3 & 0.075 & 0.050 & 0.232 & 0.254 &  9.34 & 25.07 & 24.92 & -0.15 \\
D15 & 500 & 0.2 & 0.045 & 5 & 3 & 0.020 & 0.100 & 0.348 & 0.344 & -1.31 & 23.29 & 23.94 &  0.65 \\
D16 & 500 & 0.2 & 0.045 & 5 & 5 & 0.060 & 0.050 & 0.474 & 0.463 & -2.31 & 22.60 & 22.51 & -0.09 \\
D17 & 500 & 0.2 & 0.045 & 9 & 3 & 0.033 & 0.100 & 0.381 & 0.395 &  3.52 & 23.75 & 23.17 & -0.58 \\
\midrule
D18 & 500 & 5.0 & 0.009 & 2 & 3 & 0.225 & 0.100 & 3.500 & 3.498 & -0.06 & 22.07 & 22.28 &  0.21 \\
D19 & 500 & 5.0 & 0.009 & 2 & 3 & 0.075 & 0.050 & 0.951 & 0.974 &  2.34 & 26.02 & 27.20 &  1.18 \\
D20 & 500 & 5.0 & 0.009 & 5 & 3 & 0.020 & 0.100 & 1.970 & 1.973 &  0.14 & 23.83 & 24.94 &  1.11 \\
D21 & 500 & 5.0 & 0.009 & 5 & 5 & 0.060 & 0.050 & 4.715 & 4.939 &  4.76 & 22.16 & 22.84 &  0.68 \\
D22 & 500 & 5.0 & 0.009 & 9 & 3 & 0.033 & 0.100 & 3.091 & 3.314 &  7.21 & 22.58 & 22.39 & -0.19 \\
\bottomrule
\end{tabular*}
\end{table*}

\begin{table*}[!tb]
\centering
\caption{Comparison of the VTLM and three-dimensional reference results for nonuniform transverse fin arrays.}
\label{tab:nonuniform_transverse_fin_arrays}
\scriptsize
\setlength{\tabcolsep}{1.5pt}
\renewcommand{\arraystretch}{1.5}
\begin{tabular*}{0.8\textwidth}{@{\extracolsep{\fill}}ccccccccccccccc@{}}
\toprule
Case & \(\mathrm{Re}\) & \(\mathrm{AR}\) & \(L^{+}\) & \(N_y\) & \(\lambda_y\) & \(\lambda_{x,1}\) & \(\lambda_{x,2}\) & \(\lambda_{x,3}\) & \(f_{\rm D}^{\rm 3D}\) & \(f_{\rm D}^{\rm VTLM}\) & \(e_f\) (\%) & \(\overline{T}_{\rm bt}^{\rm 3D}\) (\(^\circ\mathrm{C}\)) & \(\overline{T}_{\rm bt}^{\rm VTLM}\) (\(^\circ\mathrm{C}\)) & \(\Delta T_{\rm bt}\) (\(^\circ\mathrm{C}\)) \\
\midrule
NT1  & 100 & 1.0 & 0.075 & 3 & 0.1 & 0.500 & 0.500 & 0.500 & 6.755 & 6.495 & -3.85 & 24.42 & 24.89 &  0.47 \\
NT2  & 100 & 1.0 & 0.075 & 3 & 0.1 & 0.500 & 0.375 & 0.250 & 4.839 & 4.670 & -3.49 & 25.28 & 25.83 &  0.55 \\
NT3  & 100 & 1.0 & 0.075 & 3 & 0.1 & 0.500 & 0.250 & 0.500 & 3.033 & 2.971 & -2.05 & 27.25 & 27.92 &  0.67 \\
NT4  & 100 & 1.0 & 0.075 & 3 & 0.1 & 0.250 & 0.500 & 0.250 & 2.537 & 2.495 & -1.65 & 28.22 & 29.09 &  0.87 \\
\midrule
NT5  & 500 & 1.0 & 0.015 & 3 & 0.1 & 0.500 & 0.500 & 0.500 & 1.482 & 1.478 & -0.29 & 22.44 & 22.84 &  0.40 \\
NT6  & 500 & 1.0 & 0.015 & 3 & 0.1 & 0.500 & 0.375 & 0.250 & 1.122 & 1.122 &  0.02 & 22.85 & 23.41 &  0.56 \\
NT7  & 500 & 1.0 & 0.015 & 3 & 0.1 & 0.500 & 0.250 & 0.500 & 0.763 & 0.777 &  1.89 & 24.00 & 24.75 &  0.75 \\
NT8  & 500 & 1.0 & 0.015 & 3 & 0.1 & 0.250 & 0.500 & 0.250 & 0.649 & 0.666 &  2.65 & 24.41 & 25.39 &  0.98 \\
\midrule
NT9  & 100 & 5.0 & 0.045 & 3 & 0.1 & 0.500 & 0.500 & 0.500 & 16.636 & 16.514 & -0.73 & 23.01 & 23.16 &  0.15 \\
NT10 & 100 & 5.0 & 0.045 & 3 & 0.1 & 0.250 & 0.500 & 0.250 & 5.599 & 5.604 &  0.09 & 26.83 & 27.21 &  0.38 \\
NT11 & 500 & 5.0 & 0.009 & 3 & 0.1 & 0.500 & 0.500 & 0.500 & 3.512 & 3.512 & -0.02 & 22.06 & 22.28 &  0.22 \\
NT12 & 500 & 5.0 & 0.009 & 3 & 0.1 & 0.250 & 0.500 & 0.250 & 1.318 & 1.328 &  0.81 & 24.60 & 25.16 &  0.56 \\
\midrule
NT13 & 100 & 0.2 & 0.225 & 3 & 0.1 & 0.500 & 0.500 & 0.500 & 1.706 & 1.678 & -1.61 & 28.94 & 27.82 & -1.12 \\
NT14 & 100 & 0.2 & 0.225 & 3 & 0.1 & 0.500 & 0.375 & 0.250 & 1.434 & 1.425 & -0.62 & 29.55 & 27.74 & -1.81 \\
NT15 & 500 & 0.2 & 0.045 & 3 & 0.1 & 0.500 & 0.500 & 0.500 & 0.395 & 0.407 &  3.03 & 24.34 & 23.77 & -0.57 \\
NT16 & 500 & 0.2 & 0.045 & 3 & 0.1 & 0.250 & 0.500 & 0.250 & 0.266 & 0.283 &  6.44 & 24.64 & 24.87 &  0.23 \\
\bottomrule
\end{tabular*}
\end{table*}

\begin{table*}[!tb]
\centering
\caption{Comparison of the VTLM and three-dimensional reference results for nonuniform streamwise segmented fin arrays.}
\label{tab:nonuniform_streamwise_fin_arrays}
\scriptsize
\setlength{\tabcolsep}{1.5pt}
\renewcommand{\arraystretch}{1.5}
\begin{tabular*}{0.8\textwidth}{@{\extracolsep{\fill}}ccccccccccccccc@{}}
\toprule
Case & \(\mathrm{Re}\) & \(\mathrm{AR}\) & \(L^{+}\) & \(N_x\) & \(\lambda_y\) & \(\lambda_{x,1}\) & \(\lambda_{x,2}\) & \(\lambda_{x,3}\) & \(f_{\rm D}^{\rm 3D}\) & \(f_{\rm D}^{\rm VTLM}\) & \(e_f\) (\%) & \(\overline{T}_{\rm bt}^{\rm 3D}\) (\(^\circ\mathrm{C}\)) & \(\overline{T}_{\rm bt}^{\rm VTLM}\) (\(^\circ\mathrm{C}\)) & \(\Delta T_{\rm bt}\) (\(^\circ\mathrm{C}\)) \\
\midrule
NS1  & 100 & 1.0 & 0.075 & 3 & 0.2 & 0.150 & 0.150 & 0.150 & 1.685 & 1.697 & 0.68 & 31.62 & 31.66 &  0.04 \\
NS2  & 100 & 1.0 & 0.075 & 3 & 0.2 & 0.150 & 0.100 & 0.050 & 1.474 & 1.487 & 0.84 & 32.72 & 32.87 &  0.15 \\
NS3  & 100 & 1.0 & 0.075 & 3 & 0.2 & 0.150 & 0.050 & 0.150 & 1.567 & 1.575 & 0.52 & 32.11 & 32.28 &  0.17 \\
NS4  & 100 & 1.0 & 0.075 & 3 & 0.2 & 0.050 & 0.150 & 0.050 & 1.400 & 1.412 & 0.84 & 33.05 & 33.35 &  0.30 \\
NS5  & 100 & 1.0 & 0.075 & 3 & 0.2 & 0.050 & 0.100 & 0.150 & 1.475 & 1.487 & 0.81 & 32.61 & 32.87 &  0.26 \\
\midrule
NS6  & 500 & 1.0 & 0.015 & 3 & 0.2 & 0.150 & 0.150 & 0.150 & 0.434 & 0.464 & 7.13 & 25.88 & 26.42 &  0.54 \\
NS7  & 500 & 1.0 & 0.015 & 3 & 0.2 & 0.150 & 0.100 & 0.050 & 0.404 & 0.435 & 7.70 & 26.09 & 26.94 &  0.85 \\
NS8  & 500 & 1.0 & 0.015 & 3 & 0.2 & 0.150 & 0.050 & 0.150 & 0.421 & 0.451 & 7.03 & 25.81 & 26.65 &  0.84 \\
NS9  & 500 & 1.0 & 0.015 & 3 & 0.2 & 0.050 & 0.150 & 0.050 & 0.397 & 0.425 & 7.21 & 25.97 & 27.12 &  1.15 \\
NS10 & 500 & 1.0 & 0.015 & 3 & 0.2 & 0.050 & 0.100 & 0.150 & 0.407 & 0.435 & 6.88 & 25.87 & 26.91 &  1.04 \\
\midrule
NS11 & 100 & 5.0 & 0.045 & 3 & 0.2 & 0.150 & 0.150 & 0.150 & 3.316 & 3.366 & 1.53 & 26.27 & 26.57 &  0.30 \\
NS12 & 100 & 5.0 & 0.045 & 3 & 0.2 & 0.050 & 0.150 & 0.050 & 2.553 & 2.600 & 1.83 & 28.64 & 28.92 &  0.28 \\
NS13 & 500 & 5.0 & 0.009 & 3 & 0.2 & 0.150 & 0.150 & 0.150 & 0.779 & 0.799 & 2.59 & 24.20 & 24.45 &  0.25 \\
NS14 & 500 & 5.0 & 0.009 & 3 & 0.2 & 0.050 & 0.150 & 0.050 & 0.684 & 0.693 & 1.36 & 25.26 & 25.95 &  0.69 \\
\midrule
NS15 & 100 & 0.2 & 0.225 & 3 & 0.2 & 0.150 & 0.150 & 0.150 & 0.998 & 1.044 & 4.53 & 34.60 & 32.40 & -2.20 \\
NS16 & 100 & 0.2 & 0.225 & 3 & 0.2 & 0.150 & 0.100 & 0.050 & 0.948 & 0.988 & 4.21 & 34.20 & 32.17 & -2.03 \\
NS17 & 500 & 0.2 & 0.045 & 3 & 0.2 & 0.150 & 0.150 & 0.150 & 0.236 & 0.260 & 10.23 & 27.16 & 26.27 & -0.89 \\
NS18 & 500 & 0.2 & 0.045 & 3 & 0.2 & 0.150 & 0.100 & 0.050 & 0.226 & 0.253 & 12.12 & 26.52 & 25.98 & -0.54 \\
\bottomrule
\end{tabular*}
\end{table*}

\subsubsection{Uniform two-directional fin arrays}
\label{subsec:uniform_two_directional_fin_arrays}

Uniform two-directional fin arrays were examined using the configuration shown in Fig.~\ref{fig:uniform_fin_array_groups}(c). In these cases, both \(N_x\) and \(N_y\) exceed unity, and the fluid region is divided into multiple passages in the streamwise and transverse directions. The uniform two-directional fin-array cases and corresponding results are summarized in Table~\ref{tab:uniform_two_directional_ar1} for \(\mathrm{AR}=1\) and in Table~\ref{tab:uniform_two_directional_ar_effects} for \(\mathrm{AR}=0.2\) and \(5.0\).

For the \(\mathrm{AR}=1\) cases, reducing \(L^{+}\) from \(0.075\) to \(0.015\) increases the maximum hydraulic error from \(5.08\%\) to \(6.86\%\), whereas the maximum absolute bottom-temperature difference remains nearly unchanged, increasing only from \(1.27\) to \(1.28~^\circ\mathrm{C}\). The hydraulic error depends on the detailed in-plane arrangement rather than simply on the total fin number. For example, although Case D12 contains more fins than Case D11, its hydraulic error is only \(-0.34\%\), compared with \(-6.86\%\) for Case D11.

This behavior is consistent with the subdivision of the fluid region into narrower passages. As more of the hydraulic resistance is resolved through the in-plane fin boundaries, the unresolved top- and bottom-wall contribution represents a smaller fraction of the total pressure loss. Therefore, increasing the array density does not necessarily increase the error associated with the wall-friction closure.

The low- and high-aspect-ratio cases show the same general behavior. Across all two-directional arrays, the maximum absolute hydraulic error is \(9.34\%\), while the maximum absolute bottom-temperature difference is \(1.28~^\circ\mathrm{C}\). The dependence on \(L^{+}\) and \(\mathrm{AR}\) remains consistent with the preceding fin-array groups, and no additional limitation attributable specifically to two-directional subdivision is evident.

\begin{figure}[h]
\centering
\includegraphics[width=0.65\linewidth]{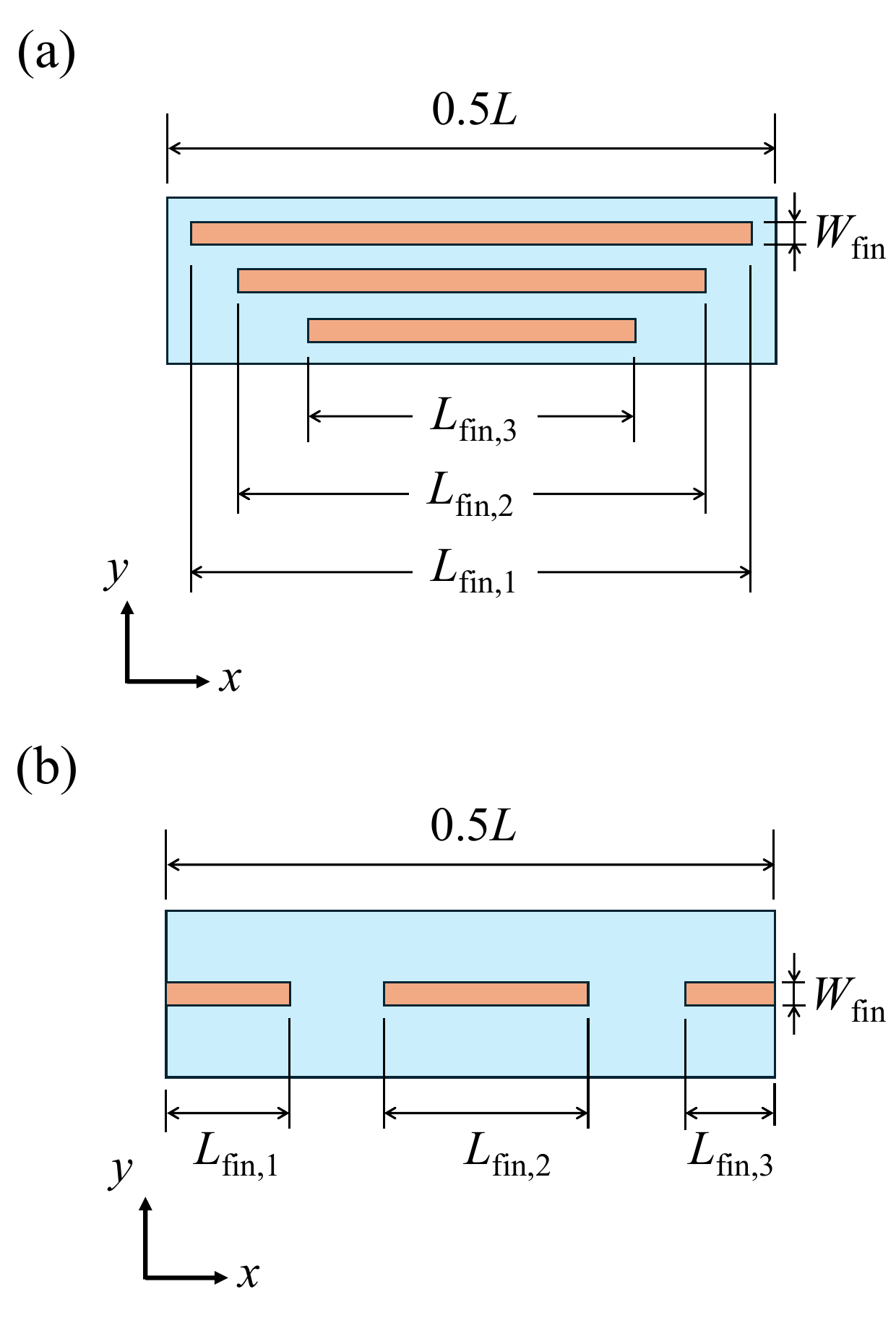}
\caption{Nonuniform fin-array configurations considered in the VTLM assessment: (a) transverse fin arrays and (b) streamwise segmented fin arrays. The streamwise fin lengths \(L_{{\rm fin},i}\) vary among fins.}
\label{fig:nonuniform_fin_array_configurations}
\end{figure}

\subsubsection{Nonuniform fin arrays}
\label{subsec:nonuniform_fin_arrays}

Fin arrays with unequal streamwise fin lengths were further examined to assess the representative fin-length treatment used in the thermal closure. The configurations considered here are shown in Fig.~\ref{fig:nonuniform_fin_array_configurations}. In the transverse arrays, the fins are centered within the heated section but have different streamwise lengths. In the streamwise segmented arrays, the upstream and downstream segments are aligned with the ends of the heated section, while the middle segment is centered. The individual streamwise fin-length ratios are denoted by \(\lambda_{x,i}=L_{{\rm fin},i}/L\), where \(i\) identifies each fin. The representative fin length defined in Eq.~\eqref{eq:equivalent_fin_length} was used to evaluate the fin-side thermal-development length for each nonuniform array.

For the nonuniform transverse arrays, \(N_y=3\) and \(\lambda_y=0.1\) were fixed, while \(\lambda_{x,1}\), \(\lambda_{x,2}\), and \(\lambda_{x,3}\) were varied. The nonuniform transverse fin-array cases and corresponding results are summarized in Table~\ref{tab:nonuniform_transverse_fin_arrays}. Cases with equal values of \(\lambda_{x,1}\), \(\lambda_{x,2}\), and \(\lambda_{x,3}\) provide uniform-array references under each operating condition.

For the nonuniform transverse arrays, redistributing the three streamwise fin lengths produces no systematic deterioration relative to the corresponding uniform references. At \(\mathrm{AR}=1\) and \(5.0\), the hydraulic errors and bottom-temperature differences remain comparable to those of the uniform arrays. Larger discrepancies occur at \(\mathrm{AR}=0.2\), including a bottom-temperature difference of \(-1.81~^\circ\mathrm{C}\) in Case NT14 and a hydraulic error of \(6.44\%\) in Case NT16. However, the discrepancies do not increase systematically as the differences among the fin lengths become larger. These results support the use of the representative fin length for transverse variations in streamwise fin length.

The nonuniform streamwise segmented fin-array cases and corresponding results are summarized in Table~\ref{tab:nonuniform_streamwise_fin_arrays}. These arrays show the same tendency. At \(L^{+}=0.075\), redistributing the segment lengths has little effect, whereas at smaller \(L^{+}\), the hydraulic discrepancy increases for both the uniform and nonuniform cases. The largest hydraulic error of \(12.12\%\) occurs in shallow-channel Case NS18, while the largest absolute bottom-temperature difference of \(2.20~^\circ\mathrm{C}\) occurs in the uniform reference Case NS15. Therefore, the reduced accuracy is associated primarily with the aspect ratio and flow development rather than with the unequal segment lengths. Overall, no additional limitation attributable specifically to fin-length nonuniformity is identified within the tested range.

\subsubsection{Model applicability}
\label{subsec:model_applicability}

The preceding numerical comparisons show that the hydraulic accuracy of the VTLM is governed primarily by the dimensionless flow-development length \(L^{+}\). For the \(\mathrm{AR}=1\) configurations, the hydraulic discrepancies generally increase as \(L^{+}\) approaches \(0.015\), whereas the detailed fin-array arrangement has a secondary influence. Fin-length nonuniformity does not introduce an additional limitation, since its error levels remain comparable to those of the corresponding uniform arrays.

The aspect ratio modifies this behavior through the relative contribution of wall friction. In shallow channels, the unresolved top- and bottom-wall contributions represent a larger portion of the total hydraulic resistance, making the prediction more sensitive to the equivalent straight-channel treatment. By contrast, this contribution becomes less dominant as the aspect ratio increases. Dense fin arrays may also reduce the sensitivity by partitioning the fluid region into narrower passages, for which the relative contribution of the unresolved wall friction is smaller. Based on these observations, a conservative lower-bound guideline for the straight-channel fin-array cases is expressed as
\begin{equation}
L^{+}>0.015\max\left(1,\frac{D_{\rm h,ref}}{D_{\rm h}}\right),
\label{eq:model_applicability}
\end{equation}
where \(D_{\rm h}\) is the hydraulic diameter of the channel under consideration and \(D_{\rm h,ref}\) is the hydraulic diameter of the \(\mathrm{AR}=1\) reference channel. The hydraulic diameters in Eq.~\eqref{eq:model_applicability} are evaluated from the original channel cross-section rather than from the equivalent geometry used in the wall-friction closure. For the present geometry, Eq.~\eqref{eq:model_applicability} gives lower bounds of \(L^{+}>0.045\) at \(\mathrm{AR}=0.2\) and \(L^{+}>0.015\) at \(\mathrm{AR}=1\).

Eq.~\eqref{eq:model_applicability} should be interpreted as a screening guideline for straight channels containing the fin-array configurations examined in this study, rather than as a strict validity boundary. Hydraulic errors may approach or slightly exceed \(10\%\) near the estimated lower limit, particularly in low-aspect-ratio channels with sparse or strongly localized blockage. The absolute bottom-temperature differences are generally less sensitive to \(L^{+}\), although larger deviations may occur when a shallow channel, a near-limit flow-development condition, and a large localized transverse blockage are combined. Accordingly, additional assessment may be necessary for configurations near the lower bound or beyond the geometric and operating ranges examined in this study.

\subsection{Fin arrays in an expansion--contraction channel}
\label{subsec:expansion_contraction_channel}

\begin{figure}[h]
\centering
\includegraphics[width=0.75\linewidth]{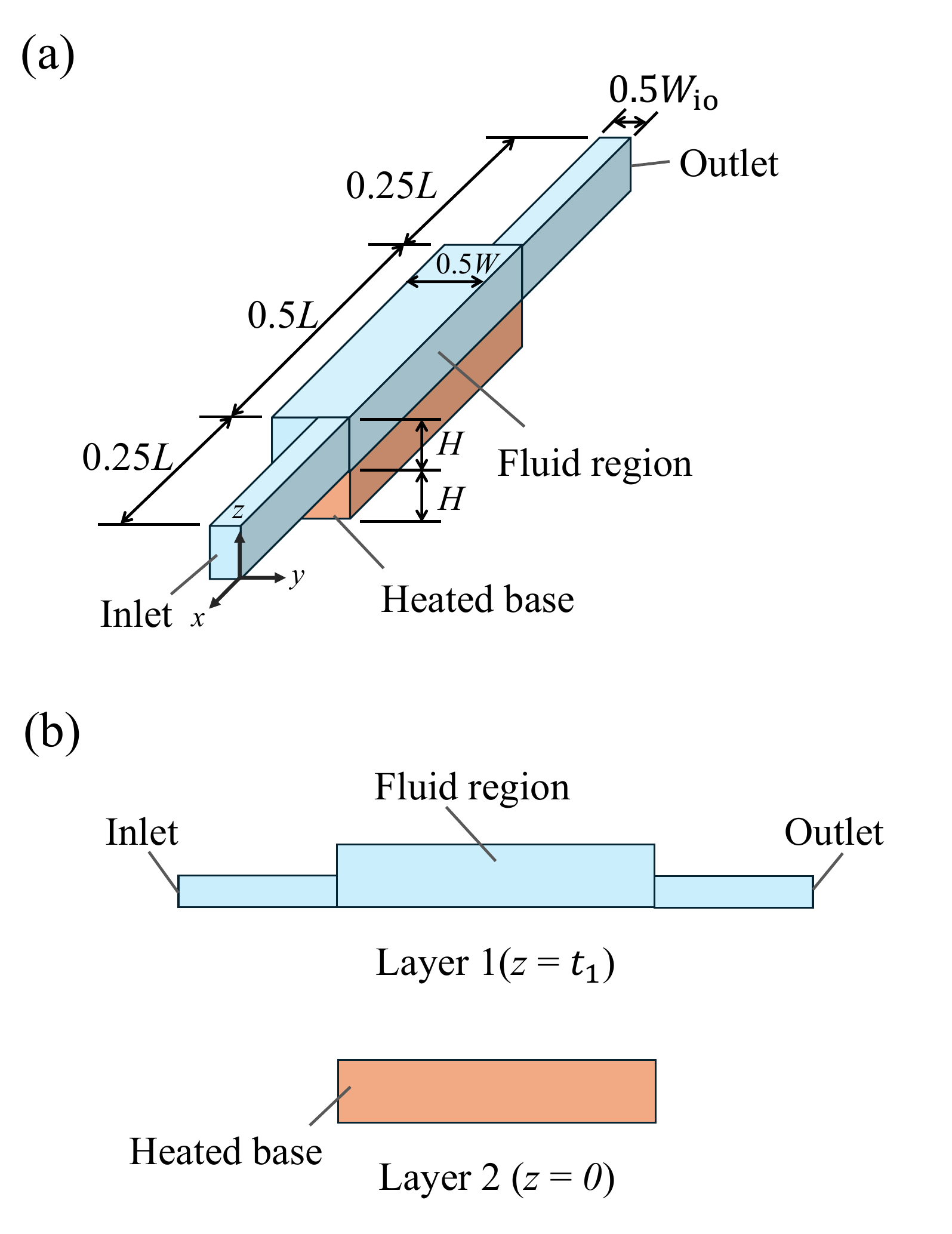}
\caption{Geometry used to assess the VTLM in an expansion--contraction channel: (a) symmetry-reduced three-dimensional reference domain and (b) corresponding Layer~1 and Layer~2 domains.}
\label{fig:expansion_contraction_domain}
\end{figure}

To evaluate the VTLM beyond straight channels of uniform width, the configuration shown in Fig.~\ref{fig:expansion_contraction_domain} was considered. The domain comprises a central heated section of length \(0.5L\), preceded and followed by inlet and outlet sections of length \(0.25L\), respectively. The heated section has a width of \(W\), whereas the inlet and outlet have a common width \(W_{\rm io}\). Owing to symmetry about the channel centerline, only half of the domain was modeled. The expansion ratio is defined as \(\mathrm{ER}=W/W_{\rm io}\). The channel height remains unchanged throughout the domain, so variations in \(\mathrm{ER}\) arise solely from the lateral change in channel width.

\begin{table*}[!tb]
\centering
\caption{Comparison of the VTLM and three-dimensional reference results for expansion--contraction channels without fins.}
\label{tab:expansion_contraction_baseline}
\scriptsize
\setlength{\tabcolsep}{3.5pt}
\renewcommand{\arraystretch}{1.5}
\begin{tabular*}{0.8\textwidth}{@{\extracolsep{\fill}}cccccccccc@{}}
\toprule
Case & \(\mathrm{Re}\) & \(\mathrm{ER}\) & \(L^{+}\) & \(f_{\rm D}^{\rm 3D}\) & \(f_{\rm D}^{\rm VTLM}\) & \(e_f\) (\%) & \(\overline{T}_{\rm bt}^{\rm 3D}\) (\(^\circ\mathrm{C}\)) & \(\overline{T}_{\rm bt}^{\rm VTLM}\) (\(^\circ\mathrm{C}\)) & \(\Delta T_{\rm bt}\) (\(^\circ\mathrm{C}\)) \\
\midrule
E1 & 100 & 2 & 0.056 & 0.503 & 0.552 &  9.65 & 46.53 & 42.91 & -3.62 \\
E2 & 100 & 3 & 0.050 & 0.500 & 0.533 &  6.57 & 42.78 & 39.44 & -3.34 \\
E3 & 100 & 4 & 0.047 & 0.501 & 0.529 &  5.44 & 41.45 & 37.96 & -3.49 \\
\midrule
E4 & 300 & 2 & 0.019 & 0.228 & 0.262 & 15.17 & 38.26 & 35.99 & -2.27 \\
E5 & 300 & 3 & 0.017 & 0.226 & 0.259 & 14.68 & 36.30 & 35.55 & -0.75 \\
E6 & 300 & 4 & 0.016 & 0.227 & 0.265 & 16.96 & 35.29 & 32.94 & -2.35 \\
\midrule
E7 & 500 & 2 & 0.011 & 0.164 & 0.195 & 18.58 & 34.85 & 33.71 & -1.14 \\
E8 & 500 & 3 & 0.010 & 0.162 & 0.192 & 17.99 & 33.18 & 33.67 &  0.49 \\
E9 & 500 & 4 & 0.009 & 0.162 & 0.189 & 16.43 & 32.35 & 33.34 &  0.99 \\
\bottomrule
\end{tabular*}
\end{table*}

\begin{table*}[!tb]
\centering
\caption{Comparison of the VTLM and three-dimensional reference results for single-fin expansion--contraction channels at \(\mathrm{ER}=2\).}
\label{tab:finned_expansion_contraction_single}
\scriptsize
\setlength{\tabcolsep}{3.5pt}
\renewcommand{\arraystretch}{1.5}
\begin{tabular*}{0.8\textwidth}{@{\extracolsep{\fill}}cccccccccccc@{}}
\toprule
Case & \(\mathrm{Re}\) & \(\mathrm{ER}\) & \(L^{+}\) & \(\lambda_x\) & \(\lambda_y\) & \(f_{\rm D}^{\rm 3D}\) & \(f_{\rm D}^{\rm VTLM}\) & \(e_f\) (\%) & \(\overline{T}_{\rm bt}^{\rm 3D}\) (\(^\circ\mathrm{C}\)) & \(\overline{T}_{\rm bt}^{\rm VTLM}\) (\(^\circ\mathrm{C}\)) & \(\Delta T_{\rm bt}\) (\(^\circ\mathrm{C}\)) \\
\midrule
EF1  & 100 & 2 & 0.056 & 0.50 & 0.05 & 0.664 & 0.681 &  2.54 & 33.04 & 32.69 & -0.35 \\
EF2  & 100 & 2 & 0.056 & 0.50 & 0.10 & 0.734 & 0.747 &  1.80 & 32.52 & 31.94 & -0.58 \\
EF3  & 100 & 2 & 0.056 & 0.50 & 0.20 & 1.009 & 1.001 & -0.72 & 31.55 & 31.46 & -0.09 \\
EF4  & 100 & 2 & 0.056 & 0.30 & 0.05 & 0.583 & 0.608 &  4.32 & 35.32 & 34.99 & -0.33 \\
EF5  & 100 & 2 & 0.056 & 0.10 & 0.05 & 0.550 & 0.581 &  5.50 & 38.86 & 37.85 & -1.01 \\
\midrule
EF6  & 500 & 2 & 0.011 & 0.50 & 0.05 & 0.223 & 0.241 &  8.23 & 26.26 & 27.03 &  0.77 \\
EF7  & 500 & 2 & 0.011 & 0.50 & 0.10 & 0.266 & 0.280 &  5.17 & 25.45 & 26.54 &  1.09 \\
EF8  & 500 & 2 & 0.011 & 0.50 & 0.20 & 0.483 & 0.475 & -1.78 & 24.46 & 25.38 &  0.92 \\
EF9  & 500 & 2 & 0.011 & 0.30 & 0.05 & 0.191 & 0.211 & 10.31 & 27.24 & 28.33 &  1.09 \\
EF10 & 500 & 2 & 0.011 & 0.10 & 0.05 & 0.182 & 0.206 & 12.90 & 28.98 & 29.93 &  0.95 \\
\bottomrule
\end{tabular*}
\end{table*}

\begin{table*}[!tb]
\centering
\caption{Comparison of the VTLM and three-dimensional reference results for streamwise segmented and two-directional fin arrays in the expansion--contraction channel at \(\mathrm{ER}=2\).}
\label{tab:finned_expansion_contraction_arrays}
\scriptsize
\setlength{\tabcolsep}{2.0pt}
\renewcommand{\arraystretch}{1.5}
\begin{tabular*}{0.8\textwidth}{@{\extracolsep{\fill}}cccccccccccccc@{}}
\toprule
Case & \(\mathrm{Re}\) & \(\mathrm{ER}\) & \(L^{+}\) & \(N_x\) & \(N_y\) & \(\lambda_x\) & \(\lambda_y\) & \(f_{\rm D}^{\rm 3D}\) & \(f_{\rm D}^{\rm VTLM}\) & \(e_f\) (\%) & \(\overline{T}_{\rm bt}^{\rm 3D}\) (\(^\circ\mathrm{C}\)) & \(\overline{T}_{\rm bt}^{\rm VTLM}\) (\(^\circ\mathrm{C}\)) & \(\Delta T_{\rm bt}\) (\(^\circ\mathrm{C}\)) \\
\midrule
ES1 & 100 & 2 & 0.056 & 2 & 1 & 0.18 & 0.05 & 0.601 & 0.624 &  3.67 & 34.30 & 33.98 & -0.32 \\
ES2 & 100 & 2 & 0.056 & 2 & 1 & 0.10 & 0.05 & 0.586 & 0.607 &  3.66 & 35.38 & 35.09 & -0.29 \\
ES3 & 100 & 2 & 0.056 & 2 & 1 & 0.02 & 0.05 & 0.565 & 0.590 &  4.41 & 38.86 & 38.25 & -0.61 \\
ES4 & 100 & 2 & 0.056 & 3 & 1 & 0.11 & 0.05 & 0.600 & 0.622 &  3.68 & 34.44 & 34.04 & -0.40 \\
\midrule
ES5 & 500 & 2 & 0.011 & 2 & 1 & 0.18 & 0.05 & 0.196 & 0.215 &  9.37 & 26.86 & 27.74 &  0.88 \\
ES6 & 500 & 2 & 0.011 & 2 & 1 & 0.10 & 0.05 & 0.193 & 0.213 & 10.64 & 27.39 & 28.30 &  0.91 \\
ES7 & 500 & 2 & 0.011 & 2 & 1 & 0.06 & 0.05 & 0.191 & 0.213 & 11.34 & 27.99 & 28.89 &  0.90 \\
ES8 & 500 & 2 & 0.011 & 3 & 1 & 0.11 & 0.05 & 0.196 & 0.215 &  9.41 & 26.95 & 27.78 &  0.83 \\
ES9 & 500 & 2 & 0.011 & 3 & 1 & 0.05 & 0.05 & 0.194 & 0.214 & 10.16 & 27.52 & 28.50 &  0.98 \\
\midrule
ET1  & 100 & 2 & 0.056 & 2 & 3 & 0.18 & 0.05 &  0.907 &  0.894 & -1.47 & 26.58 & 26.87 &  0.29 \\
ET2  & 100 & 2 & 0.056 & 2 & 3 & 0.10 & 0.05 &  0.770 &  0.771 &  0.10 & 28.03 & 28.36 &  0.33 \\
ET3  & 100 & 2 & 0.056 & 2 & 3 & 0.06 & 0.05 &  0.697 &  0.707 &  1.35 & 29.58 & 29.90 &  0.32 \\
ET4  & 100 & 2 & 0.056 & 2 & 3 & 0.18 & 0.25 & 13.595 & 12.965 & -4.63 & 23.69 & 23.94 &  0.25 \\
ET5  & 100 & 2 & 0.056 & 2 & 5 & 0.18 & 0.05 &  1.643 &  1.577 & -4.02 & 24.19 & 24.46 &  0.27 \\
ET6  & 100 & 2 & 0.056 & 2 & 5 & 0.06 & 0.05 &  0.971 &  0.959 & -1.19 & 26.56 & 26.99 &  0.43 \\
\midrule
ET7  & 500 & 2 & 0.011 & 2 & 3 & 0.18 & 0.05 &  0.270 &  0.281 &  3.97 & 23.12 & 23.74 &  0.62 \\
ET8  & 500 & 2 & 0.011 & 2 & 3 & 0.10 & 0.05 &  0.245 &  0.259 &  5.74 & 23.71 & 24.45 &  0.74 \\
ET9  & 500 & 2 & 0.011 & 2 & 3 & 0.06 & 0.05 &  0.230 &  0.246 &  6.94 & 24.38 & 25.25 &  0.87 \\
ET10 & 500 & 2 & 0.011 & 2 & 3 & 0.18 & 0.25 &  3.371 &  3.198 & -5.11 & 21.49 & 21.88 &  0.39 \\
ET11 & 500 & 2 & 0.011 & 2 & 5 & 0.18 & 0.05 &  0.428 &  0.428 &  0.11 & 22.01 & 22.41 &  0.40 \\
ET12 & 500 & 2 & 0.011 & 2 & 5 & 0.06 & 0.05 &  0.299 &  0.307 &  2.74 & 23.10 & 23.79 &  0.69 \\
\bottomrule
\end{tabular*}
\end{table*}

In the VTLM, the fluid region is represented in Layer~1, while the heated base is modeled in Layer~2. The material properties and boundary conditions were the same as those adopted in the straight-channel fin-array cases, except for the applied bottom heat flux. Because the heated-section width increases with \(\mathrm{ER}\), the bottom heat flux was scaled as \(q''_{\rm bt}=50/\mathrm{ER}~\mathrm{kW/m^2}\) to maintain the same total heat input as in the straight-channel baseline case. For the expansion--contraction cases, the hydraulic diameter used to evaluate \(L^{+}\) was calculated from the cross-section of the expanded heated section.

The assessment was conducted in two stages. Channels without fins were first examined at \(\mathrm{ER}=2\text{--}4\) to determine the influence of the abrupt lateral expansion and contraction independently of internal structures. Finned channels were subsequently evaluated at \(\mathrm{ER}=2\), including single-fin, streamwise segmented, and two-directional configurations. The two groups are discussed separately in the following subsections.

\subsubsection{Baseline channels without fins}
\label{subsubsec:baseline_expansion_contraction}

The fin-free channels were first examined to quantify the prediction differences caused by the abrupt expansion and contraction before introducing internal fin structures. Because no solid fin region is present in Layer~1, the fin-side closure is inactive, and only the fluid-side interlayer heat-transfer coefficient is employed. The corresponding VTLM and three-dimensional results are listed in Table~\ref{tab:expansion_contraction_baseline}. The VTLM predicts a higher friction factor than the three-dimensional model in all cases. At \(\mathrm{Re}=100\), the hydraulic error ranges from \(5.44\%\) to \(9.65\%\). The error increases to \(14.68\%\)--\(16.96\%\) at \(\mathrm{Re}=300\) and \(16.43\%\)--\(18.58\%\) at \(\mathrm{Re}=500\). The increase with Reynolds number is accompanied by a reduction in \(L^{+}\), whereas the hydraulic error does not exhibit a consistent trend with \(\mathrm{ER}\) across the three Reynolds numbers. This indicates that the hydraulic discrepancy is more closely associated with the flow-development condition than with the expansion ratio alone. The remaining difference can be attributed to flow redistribution near the width discontinuities, which are not explicitly included in the equivalent straight-channel representation.

The thermal prediction shows a different trend. The magnitude of \(\Delta T_{\rm bt}\) generally decreases with increasing Reynolds number, whereas no systematic dependence on \(\mathrm{ER}\) is observed. The largest absolute difference is \(3.62~^\circ\mathrm{C}\), obtained at \(\mathrm{Re}=100\) and \(\mathrm{ER}=2\). At higher Reynolds numbers, the discrepancy becomes smaller and may change sign as \(\mathrm{ER}\) increases, reflecting the combined effects of flow redistribution and thermal development in the expanded heated section. These fin-free results provide the reference prediction differences for the finned expansion--contraction cases examined in the following subsection. Comparisons with the corresponding fin-free cases allow the influence of the fin arrangement to be distinguished from that of the expansion--contraction geometry.

\subsubsection{Finned expansion--contraction channels}
\label{subsubsec:finned_expansion_contraction}

Fins were subsequently introduced into the expanded heated section at \(\mathrm{ER}=2\). The investigated layouts comprise a single fin, streamwise segmented arrays, and arrays subdivided in both the streamwise and transverse directions. These configurations follow the same types of in-plane fin arrangement introduced for the straight-channel cases in Fig.~\ref{fig:uniform_fin_array_groups}, but are placed within the expanded region shown in Fig.~\ref{fig:expansion_contraction_domain}. For the ES and ET cases, the fins were distributed over a centered streamwise region of length \(0.4L\) within the \(0.5L\) heated section, leaving a fin-free length of \(0.05L\) at each end. The single-fin results are summarized in Table~\ref{tab:finned_expansion_contraction_single}, while the streamwise segmented and two-directional results are listed in Table~\ref{tab:finned_expansion_contraction_arrays}. For all three configuration groups, \(\mathrm{Re}=100\) and \(500\) correspond to \(L^{+}=0.056\) and \(0.011\), respectively. The smaller \(L^{+}\) at \(\mathrm{Re}=500\) is consistent with the generally larger errors observed at this Reynolds number and motivates comparison with the straight-channel screening guideline in Eq.~\eqref{eq:model_applicability}.

For the single-fin cases, the hydraulic discrepancy depends clearly on the fraction of the expanded region occupied by the fin. At fixed \(\lambda_x=0.5\), increasing \(\lambda_y\) from \(0.05\) to \(0.2\) changes \(e_f\) from \(2.54\%\) to \(-0.72\%\) at \(\mathrm{Re}=100\) and from \(8.23\%\) to \(-1.78\%\) at \(\mathrm{Re}=500\). Conversely, reducing \(\lambda_x\) at fixed \(\lambda_y=0.05\) increases the hydraulic overprediction, with the error reaching \(12.90\%\) at \(\mathrm{Re}=500\). The generally larger hydraulic errors at \(\mathrm{Re}=500\) also indicate an influence of the flow-development condition. This behavior is consistent with the fin-induced wall resistance accounting for a smaller fraction of the total pressure loss as the fin becomes shorter, thereby increasing the relative influence of the expansion and contraction losses that are not explicitly included in the closure model. No consistent dependence of the bottom-temperature difference on \(\lambda_x\) or \(\lambda_y\) is observed. At \(\mathrm{Re}=100\), \(\Delta T_{\rm bt}\) ranges from \(-1.01\) to \(-0.09~^\circ\mathrm{C}\), whereas at \(\mathrm{Re}=500\), it ranges from \(0.77\) to \(1.09~^\circ\mathrm{C}\). The maximum absolute bottom-temperature difference among the single-fin cases is \(1.09~^\circ\mathrm{C}\).

For the streamwise segmented arrays, the increase in hydraulic discrepancy between the two Reynolds numbers is more pronounced than the variation caused by the segment geometry. The hydraulic error increases from \(3.66\%\text{--}4.41\%\) at \(\mathrm{Re}=100\) to \(9.37\%\text{--}11.34\%\) at \(\mathrm{Re}=500\). By contrast, the absolute bottom-temperature differences remain within \(0.61~^\circ\mathrm{C}\) at \(\mathrm{Re}=100\) and \(0.98~^\circ\mathrm{C}\) at \(\mathrm{Re}=500\). Cases with \(N_x=2\) and \(3\) give similar results at comparable streamwise size ratios, indicating that additional segmentation does not introduce a distinct deterioration beyond that associated with the smaller \(L^{+}\).

The two-directional arrays exhibit a stronger dependence of the hydraulic prediction on the detailed transverse arrangement. At \(\mathrm{Re}=100\), the maximum absolute hydraulic error is \(4.63\%\), while the absolute bottom-temperature differences remain within \(0.43~^\circ\mathrm{C}\). At \(\mathrm{Re}=500\), the maximum absolute hydraulic error increases to \(6.94\%\), whereas the maximum absolute bottom-temperature difference is \(0.87~^\circ\mathrm{C}\). For \(N_y=3\) and \(\lambda_y=0.05\), increasing \(\lambda_x\) from \(0.06\) to \(0.18\) changes \(e_f\) from \(1.35\%\) to \(-1.47\%\) at \(\mathrm{Re}=100\) and from \(6.94\%\) to \(3.97\%\) at \(\mathrm{Re}=500\). At fixed \(N_y=3\) and \(\lambda_x=0.18\), increasing \(\lambda_y\) from \(0.05\) to \(0.25\) shifts the hydraulic prediction toward underprediction at both Reynolds numbers, whereas the bottom-temperature differences remain comparatively small.

Across the finned expansion--contraction cases, the maximum absolute hydraulic error is \(12.90\%\), while the maximum absolute bottom-temperature difference is \(1.09~^\circ\mathrm{C}\). The hydraulic error levels follow the distinction implied by the straight-channel screening guideline. At \(L^{+}=0.056\), which satisfies Eq.~\eqref{eq:model_applicability}, the maximum absolute hydraulic error is \(5.50\%\), whereas at \(L^{+}=0.011\), below the lower bound, it increases to \(12.90\%\). In contrast, the maximum absolute bottom-temperature differences are \(1.01~^\circ\mathrm{C}\) and \(1.09~^\circ\mathrm{C}\) at \(L^{+}=0.056\) and \(0.011\), respectively, indicating substantially weaker sensitivity of the thermal prediction to the flow-development condition within the finned expansion--contraction cases. The hydraulic errors remain below those of the corresponding fin-free channels at both Reynolds numbers, indicating that the internal fins do not amplify the discrepancy associated with the abrupt width changes. As the fin obstruction increases, fin-induced friction accounts for a larger fraction of the total pressure loss, reducing the relative contribution of the unresolved expansion and contraction losses. These results suggest that the straight-channel screening guideline remains informative primarily for the hydraulic prediction of the finned channel at \(\mathrm{ER}=2\), although they do not establish its general applicability to arbitrary nonuniform-width passages.

\subsection{Overall thermal-resistance assessment}
\label{subsec:thermal_resistance_assessment}

To provide an overall assessment of the thermal prediction, the thermal resistance was evaluated for all assessment cases as
\begin{equation}
R_{\rm th}
=
\frac{\overline{T}_{\rm bt}-T_{\rm in}}
{Q_{\rm in}},
\label{eq:thermal_resistance}
\end{equation}
where \(Q_{\rm in}\) is the total heat input applied to the heated bottom surface. The relative difference in thermal resistance between the VTLM and the three-dimensional reference model was then defined as
\begin{equation}
e_R
=
\frac{
R_{\rm th}^{\rm VTLM}
-
R_{\rm th}^{\rm 3D}
}{
R_{\rm th}^{\rm 3D}
}
\times 100\%.
\label{eq:thermal_resistance_error}
\end{equation}

The baseline straight-channel cases used to calibrate the representative-height closure were excluded from this assessment. Because the VTLM and three-dimensional reference model are subjected to the same heat input for each case, Eq.~\eqref{eq:thermal_resistance_error} can also be written as
\begin{equation}
e_R
=
\frac{
\Delta T_{\rm bt}
}{
\overline{T}_{\rm bt}^{\rm 3D}
-
T_{\rm in}
}
\times 100\%.
\label{eq:thermal_resistance_error_temperature}
\end{equation}

\begin{figure}[h]
\centering
\includegraphics[width=0.90\linewidth]{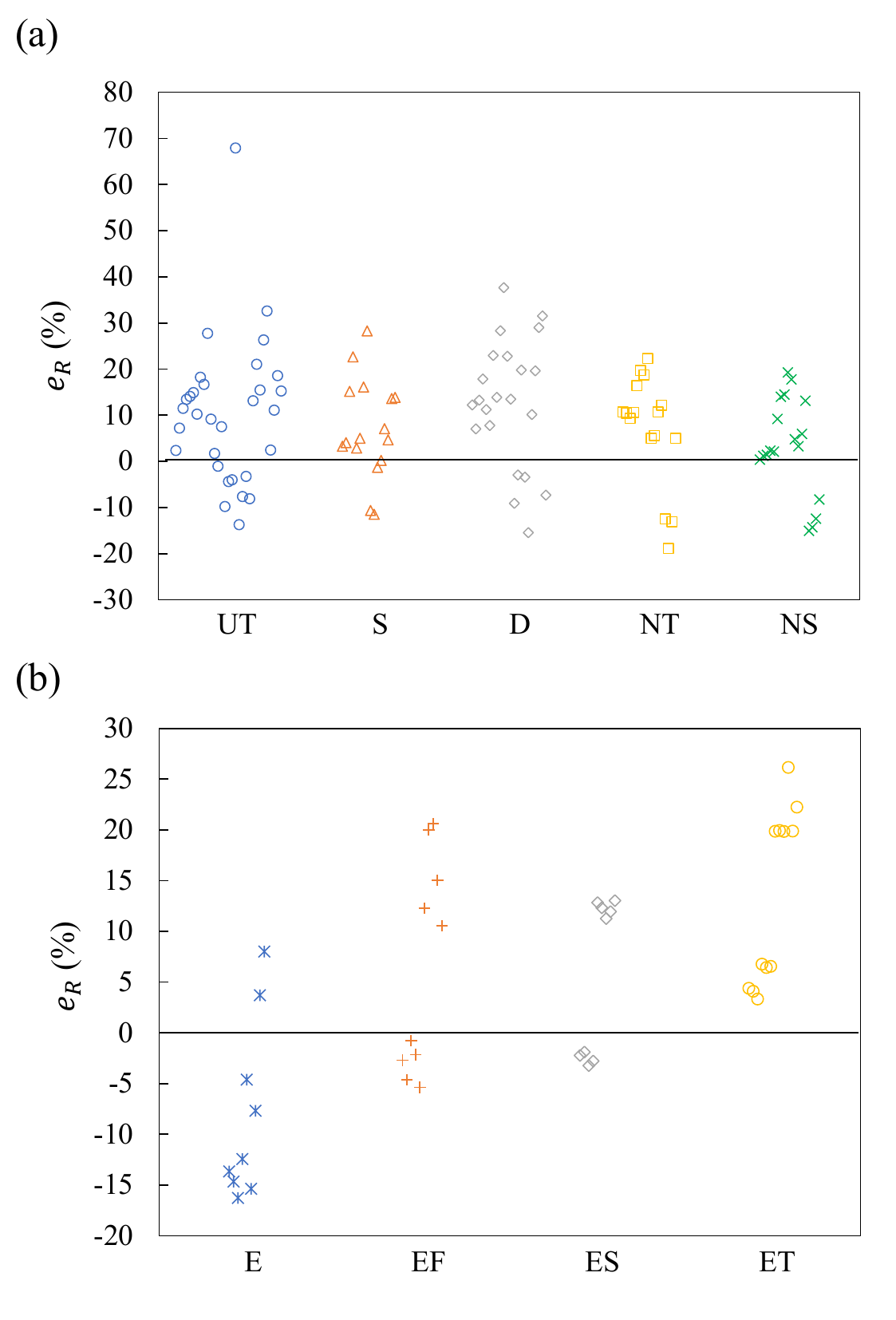}
\caption{Thermal-resistance relative differences for the assessment cases: (a) straight-channel configurations and (b) expansion--contraction configurations. Each marker represents one case. Markers within each case family are slightly offset in the horizontal direction to avoid overlap, and the horizontal offset has no physical meaning.}
\label{fig:thermal_resistance_summary}
\end{figure}

This relation shows that the thermal-resistance relative difference depends not only on the absolute bottom-temperature difference but also on the reference temperature rise. Fig.~\ref{fig:thermal_resistance_summary} summarizes the thermal-resistance relative differences for all assessment cases. Fig.~\ref{fig:thermal_resistance_summary}(a) presents the straight-channel configurations, including the uniform and nonuniform fin-array groups, whereas Fig.~\ref{fig:thermal_resistance_summary}(b) presents the fin-free and finned expansion--contraction configurations. For the straight-channel configurations, cases satisfying the screening guideline in Eq.~\eqref{eq:model_applicability} generally exhibit a more concentrated distribution of \(e_R\), whereas cases near or below the lower bound show a wider spread. The same tendency is particularly evident for the finned expansion--contraction configurations. Among the EF, ES, and ET cases, the maximum magnitude of \(e_R\) increases from approximately \(6.8\%\) to \(26.2\%\) as \(L^{+}\) decreases from \(0.056\) to \(0.011\), with the latter falling below the straight-channel lower bound. This behavior indicates that the flow-development condition identified from the preceding hydraulic assessment also provides a useful first-stage indication of thermal-resistance accuracy for the examined finned configurations.

The thermal-resistance relative difference does not depend on \(L^{+}\) alone. As indicated by Eq.~\eqref{eq:thermal_resistance_error_temperature}, the magnitude of \(e_R\) can also become large when the reference temperature rise is small. In such cases, even a small \(\Delta T_{\rm bt}\) can correspond to a comparatively large thermal-resistance relative difference. For example, Case ET10 has \(\Delta T_{\rm bt}=0.39~^\circ\mathrm{C}\) and a reference temperature rise of only \(1.49~^\circ\mathrm{C}\), resulting in \(e_R\approx26\%\). By contrast, Case UT18 exhibits a much larger \(\Delta T_{\rm bt}=4.21~^\circ\mathrm{C}\), while the thermal-resistance relative difference is also pronounced. In this case, the large thermal-resistance relative difference is mainly associated with the relatively large \(\Delta T_{\rm bt}\), rather than with a small reference temperature rise.

Accordingly, Eq.~\eqref{eq:model_applicability} remains useful as a conservative first-stage screening guideline for the overall VTLM prediction accuracy. Its strongest relevance is to the hydraulic prediction, while the thermal-resistance results indicate that it also provides a useful indication of thermal accuracy for the examined finned configurations. It should not be interpreted as a strict thermal-resistance validity boundary.

\section{Conclusion}
\label{Conclusion}

A versatile two-layer model (VTLM) was developed for thermohydraulic prediction of microchannel heat sinks under steady laminar flow. The model represents the three-dimensional conjugate heat-transfer problem using two coupled planar fields while retaining the in-plane geometry of the fluid passages and fins. Effects associated with the unresolved channel-height direction are incorporated through closures for wall friction, fluid-side heat transfer, and fin-side effective heat transfer. These closures are evaluated from the representative channel geometry, flow-development condition, and fin configuration, enabling the model to accommodate the range of channel geometries and fin arrangements examined in this study.

The three-dimensional reference model was first compared with benchmark solutions for developing laminar flow. The representative-height closure was then calibrated using baseline straight-channel conjugate heat-transfer cases spanning a broad range of channel aspect ratios. The VTLM was also compared with the existing two-layer models of Yan et al.~\cite{yan2019topology} and Zeng and Lee~\cite{zeng2019topology}. Across these baseline cases, the VTLM showed improved overall agreement with the three-dimensional reference results, highlighting the benefit of explicitly accounting for unresolved wall friction and adapting the hydraulic and thermal closures to the channel geometry and development conditions. With the closure parameters subsequently fixed, the VTLM was independently assessed using uniform transverse, streamwise segmented, two-directional, and nonuniform fin arrays. These comparisons showed that the formulation can accommodate variations in channel geometry, fin arrangement, and operating conditions without case-by-case adjustment of the closure parameters. The results further indicated that the VTLM accuracy is governed primarily by the flow-development condition and channel aspect ratio, while no distinct additional limitations associated with streamwise segmentation or fin-length nonuniformity were identified. Based on the combined straight-channel fin-array results, \(L^{+}>0.015\,\max(1,D_{\rm h,ref}/D_{\rm h})\) was proposed as a conservative screening guideline for the configurations examined in this study. The thermal-resistance assessment further showed that \(e_R\) depends on both the flow-development condition and the reference temperature rise, so the proposed guideline should not be interpreted as a strict thermal-resistance validity boundary.

The assessment was extended to expansion--contraction channels to examine the applicability of the VTLM to passages with abrupt width changes. Fin-free cases were first considered to isolate the effects of flow redistribution and local pressure losses associated with the expansion and contraction. Fins were then introduced into the expanded heated section. The comparisons showed that the wall-friction closure based on the equivalent straight-channel representation does not fully capture the local hydraulic effects caused by the width discontinuities. Nevertheless, the hydraulic discrepancies of the finned cases remained below those of the corresponding fin-free channels. This behavior is consistent with an increased contribution of the fin-induced friction captured by the VTLM to the total pressure loss. These results support the applicability of the model to the examined finned expansion--contraction configurations at \(\mathrm{ER}=2\) and suggest that the straight-channel screening guideline remains informative for these cases. Overall, the results support the VTLM as a reduced-dimensional framework for thermohydraulic analysis of the microchannel heat-sink configurations examined in this study.

Future work will extend the VTLM to more complex channel layouts, including U-shaped channels and passages with multiple bends~\cite{mandel2018modeling}, and further refine the thermal-side closure to improve thermal-resistance prediction accuracy over wider operating and geometric ranges. The extended model can then be integrated into a topology optimization framework for efficient microchannel heat-sink design.


\section*{CRediT authorship contribution statement}
\textbf{Chih-Hsiang Chen:} Writing -- original draft, Conceptualization, Methodology, Validation, Formal analysis, Investigation, Data curation, Visualization. \textbf{Kentaro Yaji:} Writing-- review \& editing, Conceptualization, Methodology, Software, Supervision, Resources, Project administration, Funding acquisition.

\section*{Acknowledgement}
This work was supported by JSPS KAKENHI (Grant No. 23K26018).

\appendix
\section{Derivation of the fin-side effective heat transfer coefficient}
\label{app:fin_side_htc_derivation}

The fin-side effective heat-transfer coefficient is derived by representing the collection of fins as an equivalent one-dimensional fin of uniform cross section. Let \(z\) denote the coordinate measured from the fin base, with \(z=0\) at the fin-base junction represented by Layer~2 and \(z=H\) at the fin tip. Under the uniform-extrusion assumption, the total cross-sectional area normal to the \(z\)-direction and the total wetted perimeter are \(A_{\rm fin}^{\rm tot}\) and \(P_{\rm fin}^{\rm tot}\), respectively. The fluid-side heat-transfer coefficient \(h_{\rm f,fin}\) and the representative Layer~1 fluid temperature \(T_{1,\rm f}\) are assumed to be uniform along the fin height.

For a differential fin element, the balance between axial conduction and lateral heat transfer gives
\begin{equation}
\frac{d}{dz}
\left(
k_{\rm s}A_{\rm fin}^{\rm tot}
\frac{dT_{\rm fin}}{dz}
\right)
-
h_{\rm f,fin}P_{\rm fin}^{\rm tot}
\left(
T_{\rm fin}-T_{1,\rm f}
\right)
=
0.
\label{eq:app_fin_energy_balance}
\end{equation}

Because \(k_{\rm s}\) and \(A_{\rm fin}^{\rm tot}\) are constant, the excess temperature is introduced as $\vartheta(z)=T_{\rm fin}(z)-T_{1,\rm f}$. Eq.~\eqref{eq:app_fin_energy_balance} then reduces to
\begin{equation}
\frac{d^{2}\vartheta}{dz^{2}}
-
m^{2}\vartheta
=
0,
\qquad
m
=
\sqrt{
\frac{
h_{\rm f,fin}P_{\rm fin}^{\rm tot}
}{
k_{\rm s}A_{\rm fin}^{\rm tot}
}
}.
\label{eq:app_fin_governing_equation}
\end{equation}

The fin-base temperature is given by the Layer~2 temperature, \(T_{\rm fin}(0)=T_2\), and the fin tip is treated as adiabatic. The corresponding boundary conditions are given as follows:
\begin{equation}
\vartheta(0)
=
\vartheta_{\rm b}
=
T_2-T_{1,\rm f},
\qquad
\left.
\frac{d\vartheta}{dz}
\right|_{z=H}
=
0.
\label{eq:app_fin_boundary_conditions}
\end{equation}

Solving Eq.~\eqref{eq:app_fin_governing_equation} subject to Eq.~\eqref{eq:app_fin_boundary_conditions} gives
\begin{equation}
\vartheta(z)
=
\vartheta_{\rm b}
\frac{
\cosh\left[m(H-z)\right]
}{
\cosh(mH)
}.
\label{eq:app_fin_temperature_solution}
\end{equation}

The axial conductive heat flow crossing the fin section at position \(z\) is obtained from Fourier's law:
\begin{equation}
Q_{\rm ax}(z)
=
-k_{\rm s}A_{\rm fin}^{\rm tot}
\frac{dT_{\rm fin}}{dz}
=
k_{\rm s}A_{\rm fin}^{\rm tot}
m\vartheta_{\rm b}
\frac{
\sinh\left[m(H-z)\right]
}{
\cosh(mH)
}.
\label{eq:app_fin_axial_heat_flow}
\end{equation}

At the representative Layer~1 height \(z=t_1\), the dimensionless parameters are defined as
\begin{equation}
\Lambda
=
mH
=
H
\sqrt{
\frac{
h_{\rm f,fin}P_{\rm fin}^{\rm tot}
}{
k_{\rm s}A_{\rm fin}^{\rm tot}
}
},
\qquad
\zeta_1
=
m(H-t_1)
=
\Lambda
\left(
1-\frac{t_1}{H}
\right).
\label{eq:app_fin_dimensionless_parameters}
\end{equation}

The remaining axial heat flow at the representative height is denoted by \(Q_{\rm top}=Q_{\rm ax}(t_1)\). Eq.~\eqref{eq:app_fin_axial_heat_flow} therefore gives
\begin{equation}
Q_{\rm top}
=
\frac{
k_{\rm s}A_{\rm fin}^{\rm tot}
}{
H
}
\Lambda\vartheta_{\rm b}
\frac{
\sinh\zeta_1
}{
\cosh\Lambda
}.
\label{eq:app_fin_heat_flow_t1}
\end{equation}

The temperature difference used to define the fin-side effective heat-transfer coefficient is obtained from Eq.~\eqref{eq:app_fin_temperature_solution}:
\begin{align}
T_2-T_{\rm fin}(t_1)
&=
\left(T_2-T_{1,\rm f}\right)
-
\left[T_{\rm fin}(t_1)-T_{1,\rm f}\right]
\nonumber\\
&=
\vartheta_{\rm b}
\left(
1-
\frac{
\cosh\zeta_1
}{
\cosh\Lambda
}
\right)
\nonumber\\
&=
\vartheta_{\rm b}
\frac{
\cosh\Lambda-\cosh\zeta_1
}{
\cosh\Lambda
}.
\label{eq:app_fin_temperature_difference}
\end{align}

Using the definition introduced in Eq.~\eqref{eq:hs_definition},
\begin{equation}
h_{\rm s}
=
\frac{
Q_{\rm top}
}{
A_{\rm fin}^{\rm tot}
\left[
T_2-T_{\rm fin}(t_1)
\right]
},
\label{eq:app_hs_definition}
\end{equation}
and substituting Eqs.~\eqref{eq:app_fin_heat_flow_t1} and \eqref{eq:app_fin_temperature_difference} yields
\begin{equation}
h_{\rm s}
=
\frac{k_{\rm s}}{H}
\frac{
\Lambda\sinh\zeta_1
}{
\cosh\Lambda-\cosh\zeta_1
}.
\label{eq:app_hs_final}
\end{equation}

This expression corresponds to Eq.~\eqref{eq:hs_closed_form}, with \(\Lambda\) and \(\zeta_1\) defined by Eq.~\eqref{eq:fin_side_parameters}.

\nocite{*}

\end{document}